%% file: paper.tex
\documentclass[sigconf]{acmart}
   
\renewcommand\footnotetextcopyrightpermission[1]{} 
\input{header}

\hypersetup{
    colorlinks=true,
    allcolors=blue,
    linkcolor=blue,
    filecolor=magenta,      
    urlcolor=cyan,
}

\begin{document}
\begin{sloppypar}

\title{An Empirical Analysis of Deep Learning \\ for Cardinality Estimation}
\author{
Jennifer Ortiz$^\dagger$, Magdalena Balazinska$^\dagger$, Johannes Gehrke$^\ddagger$, S. Sathiya Keerthi$^+$ \\ 
University of Washington$^\dagger$, Microsoft$^\ddagger$, Criteo Research$^+$}

\begin{abstract}
We implement and evaluate deep learning for cardinality estimation by studying the accuracy, space and time trade-offs across several architectures. We find that simple deep learning models can learn cardinality estimations across a variety of datasets (reducing the error by 72\% - 98\% on average compared to PostgreSQL).  In addition, we empirically evaluate the impact of injecting cardinality estimates produced by deep learning models into the PostgreSQL optimizer. In many cases, the estimates from these models lead to better query plans across all datasets, reducing the runtimes by up to 49\% on select-project-join workloads. As promising as these models are, we also discuss and address some of the challenges of using them in practice.
\end{abstract}

\maketitle


\input{intro}
\input{learning_cardinalities}
\input{measurement_analysis}

\input{practical_considerations}
\input{related_work}

\input{conclusion}
\end{sloppypar}

\nocite{*}

\input{paper.bbl}
\end{document}

%% file: header.tex
\usepackage{graphicx}
\usepackage{balance}  
\usepackage{graphics} 
\usepackage{times}    
\usepackage{url}      
\usepackage{hyperref}

\usepackage{tabularx}
\usepackage{float}
\usepackage{color}
\usepackage{xcolor}
\usepackage{url}
\usepackage[noend]{algpseudocode}
\usepackage{algorithm}
\usepackage{verbatim}
\usepackage{mathtools}
\usepackage{caption}
\usepackage{subcaption}
\usepackage{amsmath}

\usepackage{amsthm}
\usepackage{thmtools}
\usepackage{xspace}
\usepackage{multirow}
\usepackage[capitalise]{cleveref}
\usepackage{amssymb}
\usepackage{amsmath}

\usepackage{amsmath}
\usepackage{caption}
\usepackage{subcaption}

\newcommand{\ea}{{\em et al.}\xspace}

\algdef{SE}[SUBALG]{Indent}{EndIndent}{}{\algorithmicend\ }%
\algtext*{Indent}
\algtext*{EndIndent}

%% file: intro.tex
 \section{Introduction}\label{sec:deeplearning}
 
 Query optimization is at the heart of relational database management systems (DBMSs). Given a
 SQL query, the optimizer automatically generates an efficient execution plan for that query. Even though query optimization is an old problem~\cite{Selinger:79}, it remains a challenging
 problem today: existing database management systems (DBMSs) still choose poor execution plans for many queries~\cite{Leis:15}. Cardinality estimation is the ability to estimate the number of tuples produced by a subquery. This is a key component in the query optimization process. It is especially challenging with complex queries that contain many joins, where cardinality estimation errors propagate and amplify from the leaves to the root of the query plan. One problem is that existing DBMSs make simplifying assumptions about the data (e.g., inclusion principle, uniformity or independence assumptions) when estimating the cardinality of a subquery. When these assumptions do not hold, cardinality estimation errors occur, leading to sub-optimal plan selections~\cite{Leis:15}. To accurately estimate cardinalities, optimizers must be able to capture detailed data distributions and correlations across columns. Capturing and processing this information, however, imposes space and time overheads and adds complexity.


To support cardinality estimation, DBMSs collect statistics about the data. These statistics typically take the form of histograms or samples. Because databases contain many tables with many columns, these statistics rarely capture all existing correlations. The manual process of selecting the best statistics to collect can help but requires expertise both in database systems and in the application domain.

Recently, thanks to dropping hardware costs and growing datasets available for training, \textit{deep learning} has successfully been applied to solving computationally intensive learning tasks in other domains. The advantage of these type of models comes from their ability to learn unique patterns and features of the data that are difficult to manually find or design~\cite{Goodfellow:16}.

Given this success, we ask the following fundamental question: Should we consider using deep learning for query optimization? Can a deep learning model actually learn properties about the data and learn to capture correlations that exist in the data? What is the overhead of building these models? How do these models compare to other existing machine learning techniques? In this work, we implement a variety of deep learning architectures to predict query cardinalities. Instead of relying on basic statistics and formulas to estimate cardinalities, we train a model to automatically learn important properties of the data to more accurately infer these estimates. In this paper, we seek to understand the fundamental capabilities of deep neural networks for this application domain. For this reason, we focus on the performance of basic deep learning architectures. 


Our community has recently started to consider the potential of deep learning techniques to solve database problems~\cite{Wang:16_2}. There still is, however, limited understanding of the potential and impact of these models for query optimization. Previous work has demonstrated the potential of using deep learning as a critical tool for learning indexes~\cite{Kraska:18}, improving query plans~\cite{Marcus:19}, and learning cardinalities specifically through deep set models~\cite{Kipf:19}, but we argue that that the accuracy should not be the only factor to consider when evaluating these models. We also need to consider their overheads, robustness, and impact on query plan selection. We need a systematic analysis of the benefits and limitations of various fundamental architectures.

In this experimental study, we focus on the trade-offs between the size of the model (measured by the number of trainable parameters), the time it takes to train the model, and the accuracy of the predictions. We study these trade-offs for several datasets. Our goal is to understand the overheads of these models compared to PostgreSQL's optimizer. To do this, we build several simple neural network as well as recurrent neural network models and vary the complexity by modifying the network widths and depths. We train each model separately and compare the overheads of these models to PostgreSQL and tree ensembles (based on off-the-shelf machine learning models).

To summarize, we contribute the following:

\begin{itemize}
    \item We show how deep neural networks and recurrent neural networks can be applied to the problem of cardinality estimation and describe this process in \autoref{sec:cardinalities}. 
    \item We comparatively evaluate neural networks, recurrent neural networks, tree ensembles, and PostgreSQL's optimizer on three real-world datasets in \autoref{sec:measurement_analysis}. For a known query workload, we find that, compared to PostgreSQL, simple deep learning models that are similar in space improve cardinality predictions by reducing the error by up to 98\%. These models, however, come with high training overheads. We also find that, although tree ensembles usually require a larger amount of space, they are fast to train and are more accurate than the deep learning models under certain settings. 
    \item In \autoref{sec:robustness}, we study these models in more detail by evaluating the robustness of these models with respect to query workload changes. We find that tree ensembles are more sensitive to these changes compared to the neural network and recurrent neural network models.
    \item In \autoref{sec:latents}, we visualize the embeddings from the models to understand what they are learning. 
    \item Finally, we study these models from a practical perspective in \autoref{sec:practical}. We evaluate how predictions from these models improve query plan selection. We find that these models can help the optimizer select query plans that lead from 7\% to 49\% faster query executions. 
\end{itemize}




%% file: learning_cardinalities.tex
\section{Background and Problem Statement}\label{sec:background}

Many optimizers today use histograms to estimate cardinalities. These structures can efficiently summarize the frequency distribution of one or more attributes. For single dimensions, histograms  split the data using equal-sized buckets (equi-width) or buckets with equal frequencies (equi-depth). To minimize errors, statistics about each bucket are also stored including but not limited to the number of items, average value, and mode~\cite{Cormode:12}. 

These histograms are especially relevant in cases where there are simple single query predicates. For more complex predicates, the system extracts information from these histograms in conjunction with ``magic constants'' to make predictions~\cite{Leis:15}. Optimizers typically do not build or use multidimensional histograms or sampling due to the increased overheads~\cite{Eavis:07, Wu:16}. As the estimates from these optimizers are not theoretically grounded, propagating these estimates through each intermediate result of a query plan can result in high cardinality errors, leading to sub-optimal query plans. 

In this paper, we use PostgreSQL's optimizer as representative of this class because it is a mature optimizer available in a popular open source system. 


Our goal in this paper is to apply deep learning to the cardinality estimation problem and compare the performance of this approach empirically to that of a traditional query optimizer.

We consider the following scenario: A database system is deployed at a customer's site. The customer has a database $D$ and a query workload $Q$. Both are known. We compare the following approaches:
\begin{itemize}
    \item \textbf{Traditional: } In the pre-processing phase, we build histograms on select attributes in $D$. We select those attributes following standard best practices given the workload $Q$. Simple best practices include collecting statistics for all frequently joined columns and for non-indexed columns frequently referenced as a selection predicate, particularly if the column contains very skewed data\cite{teradata:18}. We then measure the accuracy of cardinality estimates on queries in $Q$ (and queries similar to $Q$) and the overhead of creating and storing the histograms. We measure both the time it takes to build the histograms and the space that the histograms take.
    \item \textbf{Deep Neural Networks: } In the pre-processing phase, we execute all queries in $Q$ to compute their cardinalities. We use the results to train deep neural networks. We encode all queries in $Q$ into inputs for the models, and evaluate how accurately these models are able to learn the function between the input, $X$ and the cardinality value, $Y$. As above, we measure the overhead of building, storing and accuracy of cardinality estimates for queries in $Q$ and queries not in $Q$ but similar to those in $Q$. To compare different architectures, we build several models by varying the width and depth.
\end{itemize}


As a simplifying assumption, in this paper, we focus on select-project-join queries and only use single-sided range predicates as selection predicates. The join predicates consist of primary key and foreign key relationships between tables, as defined by their schema.

\section{Machine Learning-Based Cardinality Estimation}\label{sec:cardinalities}

The first contribution of this paper is to articulate how to map the cardinality estimation problem into a learning problem. We show the mapping for three types of models: neural networks, recurrent neural networks, and tree ensembles.

For ease of illustration, in this section, we use a simple running example comprising
a database $D$ with three relations, $D$ : $\{A, B, C\}$. Each relation has two attributes where relation $A$ contains $\{a_1, a_2\}$, relation $B$ has attributes $\{b_1, b_2\}$, and relation $C$ has attributes $\{c_1, c_2\}$. In this database, there are two possible join predicates. Attribute $a_2$ is a foreign key to primary key attribute $b_1$, while $b_2$ serves as a foreign key to primary key attribute $c_1$.

\subsection{Neural Networks}
\label{sec:nn}

Deep learning models are able to approximate a non-linear function, $f$~\cite{Goodfellow:16}. These models define a mapping from an input $X$ to an output $Y$, through a set of learned parameters across several layers with weights, $\theta$. Each layer contains a collection of neurons, which help express non-linearity. During training, the behavior of the inner layers are not defined by the input data $X$, instead these models must learn how to tune the weights to produce the correct output. Since there is no direct interaction between the layers and the input training data, these layers are called \textit{hidden layers}~\cite{Goodfellow:16}. 

\begin{figure}[t]
    \centering
    \includegraphics[width=.6\linewidth]{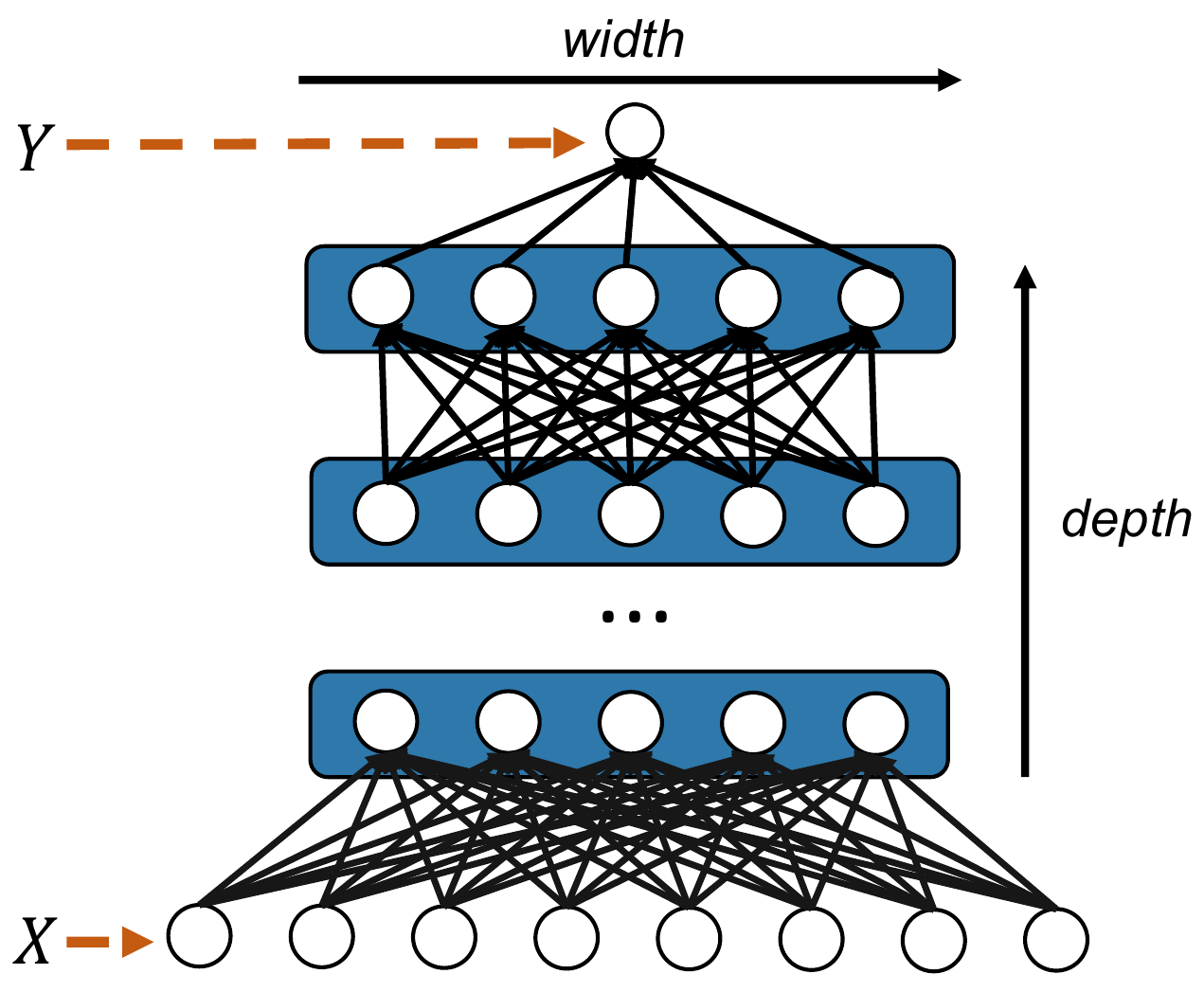}
    \caption{Illustration of a Deep Neural Network: The input consists of an input $X$. This is then fed into a network with $n$ hidden layers, which then makes a prediction for the cardinality of the query, $Y$.}
    \label{fig:nn}
\end{figure}

Training occurs through a series of steps. First, during forward propagation, a fixed-sized input $X$ is fed into the network through the input layer. This input is propagated through each layer through a series of weights~\cite{Skansi:18} until it reaches the final output, $Y$. This process is illustrated in \autoref{fig:nn}. After a forward pass, the backpropagation step then evaluates the error of the network and through gradient descent, modifies the weights to improve errors for future predictions. 

There are several architectures we could consider for the model. As shown in \autoref{fig:nn}, a neural network can have a different number of layers (depth) and a different number of hidden units in each individual layer (width). Determining the correct number of hidden units is currently an active area of research and does not have strong theoretical principles~\cite{Goodfellow:16}. Although a network with only a single wide hidden layer is capable of learning a complex function, deep networks are able to use a smaller number of training parameters to achieve the same goal. Unfortunately, deep networks are difficult to train and to optimize~\cite{Goodfellow:16}. In this work, we focus on evaluating a variety of network architectures. We focus on simple architectures comprising a small number of fully connected layers. We vary the width and the depth of the network. More complex architectures are possible~\cite{Kipf:19} and are also interesting to study. Our goal, however, is to understand the performance of basic architectures first.

Given a model, a query $q$ and a fixed dataset $D$, we define an encoding for the input, $X$. 
The input $X$ should contain enough information for the model to learn a useful mapping. There are several ways to represent a query as an input vector. The encoding determines how much information we provide the network.  In this work,  we define $X$ as a concatenation of three single dimensional vectors: $\mathcal{I}_{relations}$, $\mathcal{I}_{selpred}$, and $\mathcal{I}_{joinpred}$. To explain this encoding, we first describe how to encode selection queries.

\textbf{Modeling Selection Queries} With selection queries, the goal is to have the network learn the distribution of each column and combinations of columns for a single relation. To encode a selection query, we provide the model with information about \textit{which} relation in $D$ we are applying the selections to, along with the attribute values used in the selection predicates. We encode the relation using vector $\mathcal{I}_{relations}$ as a binary one-hot encoding. Each element in  $\mathcal{I}_{relations}$ represents a relation in $D$. If a relation is referenced in $q$, we set the designated element to 1, otherwise we set it to 0.

\begin{figure}[t]
    \centering
    \includegraphics[width=.9\linewidth]{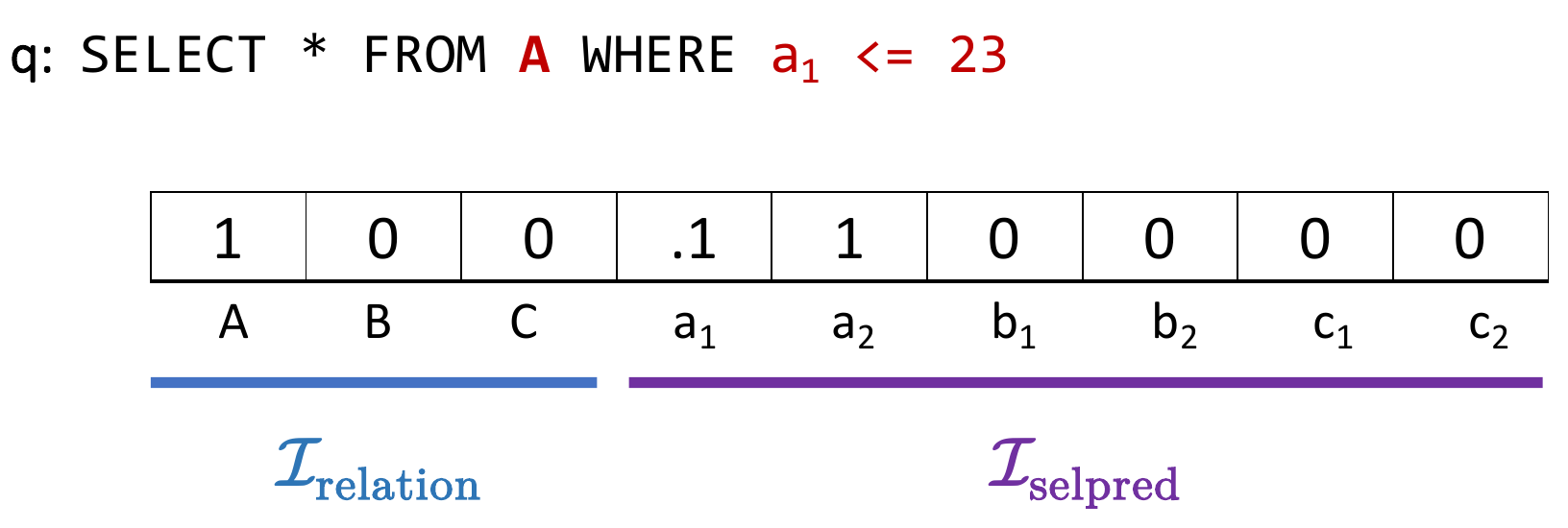}
    \caption{Query Encoding for Selections: We encode a selection by specifying the underlying relations in the query and all selection predicate values.}
    \label{fig:nn_sel}
\end{figure}

We encode the selection predicates in $q$ using vector $\mathcal{I}_{selpred}$. As described in \autoref{sec:background}, selection predicates are limited to single-sided range predicates. Each element in this vector holds the selection value for one attribute. The vector includes one element for each attribute of each relation in $D$. As an example, assume we have the following query: SELECT * FROM $A$ WHERE $a_1 \leq 23$. In this case, we set the corresponding element for $a_1$ in $\mathcal{I}_{selpred}$ as $23$. Otherwise, if there is no selection on an attribute, we set the element with the maximum value of the attribute's domain. This captures the fact that we are selecting all values for that attribute. 

Neural networks are highly sensitive to the domain of the input values. 
Having unnormalized values in the input will highly impact the error surface for gradient descent, making the model difficult to train. Instead, we encode these selection predicates as values ranging from 0 to 1, where the value represents the percentile of the attribute's active domain as shown in \autoref{fig:nn_sel}. The output of the model is also normalized to represent the percentage of tuples that remain after the selection query is applied to the relation. Using this type of normalization, instead of learning query cardinalities, we are in fact learning predicate selectivities.



\textbf{Modeling Join Queries} Introducing queries that contain both joins and selections requires the model to learn a more complex operation. Joins essentially apply a cartesian product across a set of relations followed by additional filters that correspond to the equality join and selection predicates.  We encode existing \textit{join predicates} with the vector $\mathcal{I}_{joinpred}$ using a binary one-hot encoding. As we now include joins, the output $Y$ now represents the fraction of tuples selected from the join result. Hence, once again, the model will learn the selectivity of the join operation.

\begin{figure}[t]
    \centering
    \includegraphics[width=.9\linewidth]{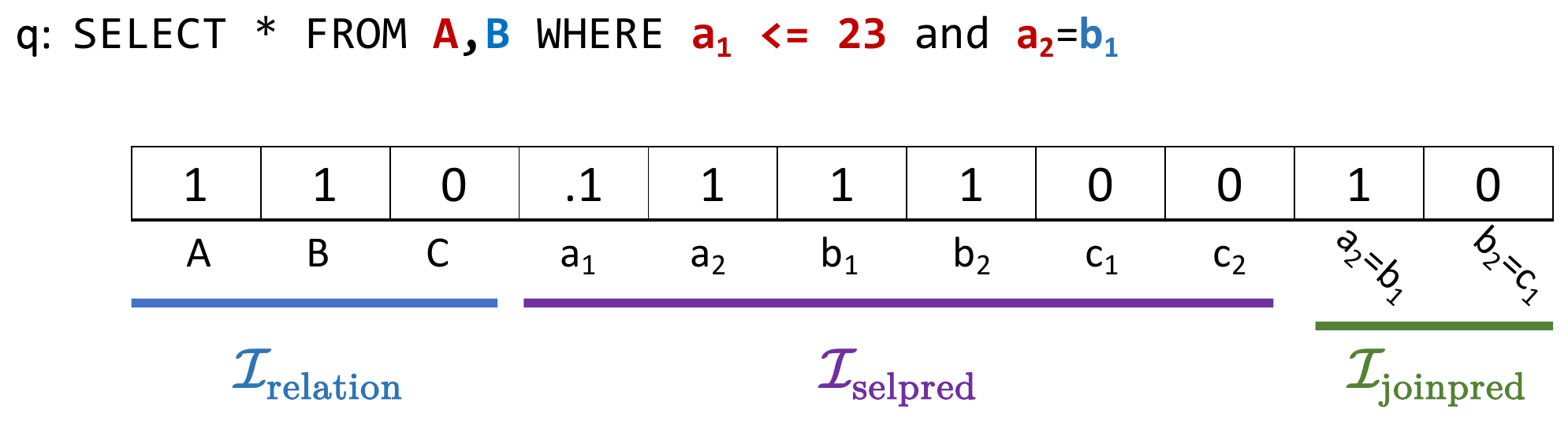}
    \caption{Query Encoding for Joins+Selections: We encode a join+selection query based on the joined relations, the selection predicate values and join predicates.}
    \label{fig:nn_join}
\end{figure}

We illustrate this encoding using an example in~\autoref{fig:nn_join}. Given the following query from our running example dataset, SELECT $*$ FROM $A, B$ WHERE $a_1 \leq 23$ and $a_2=b_1$, we show the encoding in the figure. For $\mathcal{I}_{relations}$, there are three possible elements, one for each relation in $D$. For this query, we only set the elements corresponding to relations $A$ and $B$ to 1. The vector $\mathcal{I}_{selpred}$ contains the encoding for the selection predicates. Since relation $C$ is not referenced in $q$, we set all of its attributes in $\mathcal{I}_{selpred}$ to 0. We set to $.1$ the element corresponding to attribute $a_1$, as it represents the percentile of the active domain for $a_1$. The rest of the attributes from $A$ and $B$ are set to $1$, as we are not filtering any values from these attributes. Finally, the vector $\mathcal{I}_{joinpred}$ encodes the join predicate  $a_2=b_1$ with a 1.

\subsection{Recurrent Neural Networks}

If we focus on left-deep plans, we can model queries as sequence of operations, and we can leverage that structure when learning a model~\cite{ortiz:18}. Recurrent neural networks (RNN) in particular are designed for sequential data such as time-series data or text sequences~\cite{Skansi:18}. Compared to neural networks where the input is a single vector $X$, the input to RNNs is a sequence with $t$ timesteps, $X$ = $\{ x_1, x_2, ... , x_t\}$. For each timestep $t$, the model receives two inputs: $x_t$ and $h_{t-1}$, where $h_{t-1}$ is the generated hidden state from the previous timestep~\cite{Skansi:18}. With these inputs, the model generates a hidden state for the current timestep $t$, where $h_t = f(h_{t-1}, x_t)$ and $f$ represents an activation function. Given this feedback loop, each hidden state contains traces of not only the previous timestep, but all those preceding it as well. RNNs can either have a single output $Y$ for the final timestep (a many-to-one architecture) or they can have one output for each timestep (many-to-many) where $Y = \{y_1, y_2, ..., y_t \}$. 

\begin{figure}[t]
    \centering
    \includegraphics[width=\linewidth]{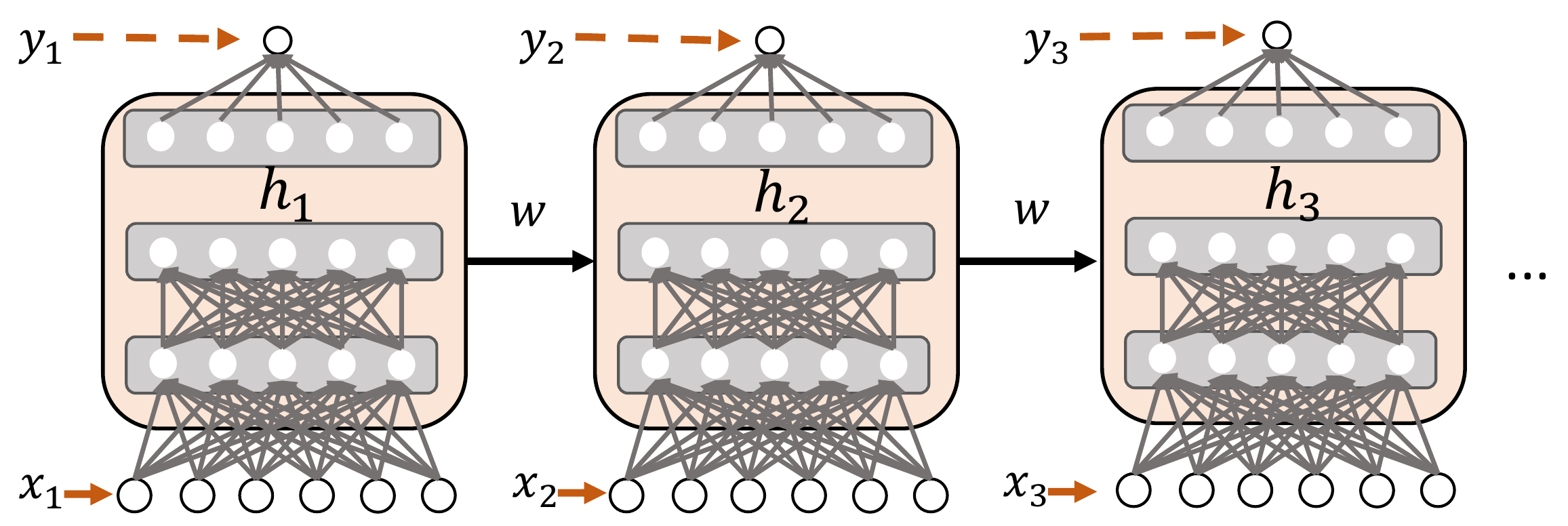}
    \caption{Illustration of a Recurrent Neural Network: The input consists of a sequence of inputs $\{x_1, x_2, ..., x_t\}$. Each input, along with the hidden state of the previous timestep, is fed into the network to make a prediction, $y_i$.}
    \label{fig:rnn}
\end{figure}

In our context, we model queries as a series of actions, where each action represents a query operation (i.e. a selection or a join) in a left-deep query plan corresponding to the query. With a sequential input, RNNs incrementally generate succinct representations for each timestep, which are known as \textit{hidden states}, and which represent a subquery. Recurrent neural networks rely on these hidden states to infer context from previous timesteps. More importantly, $h_t$ is not a manually specified feature vector, but it is the latent representation that the model learns itself. We illustrate how these hidden states are generated in \autoref{fig:rnn}. Information from each hidden state, $h_t$ is fed into the next timestep, $t+1$ through shared weights, $w$. In our context, hidden representations are useful, as they capture important details about the underlying intermediate result. The information learned at the hidden state is highly dependent on the input and output of the network. 

\begin{figure}[t]
    \centering
    \includegraphics[width=\linewidth]{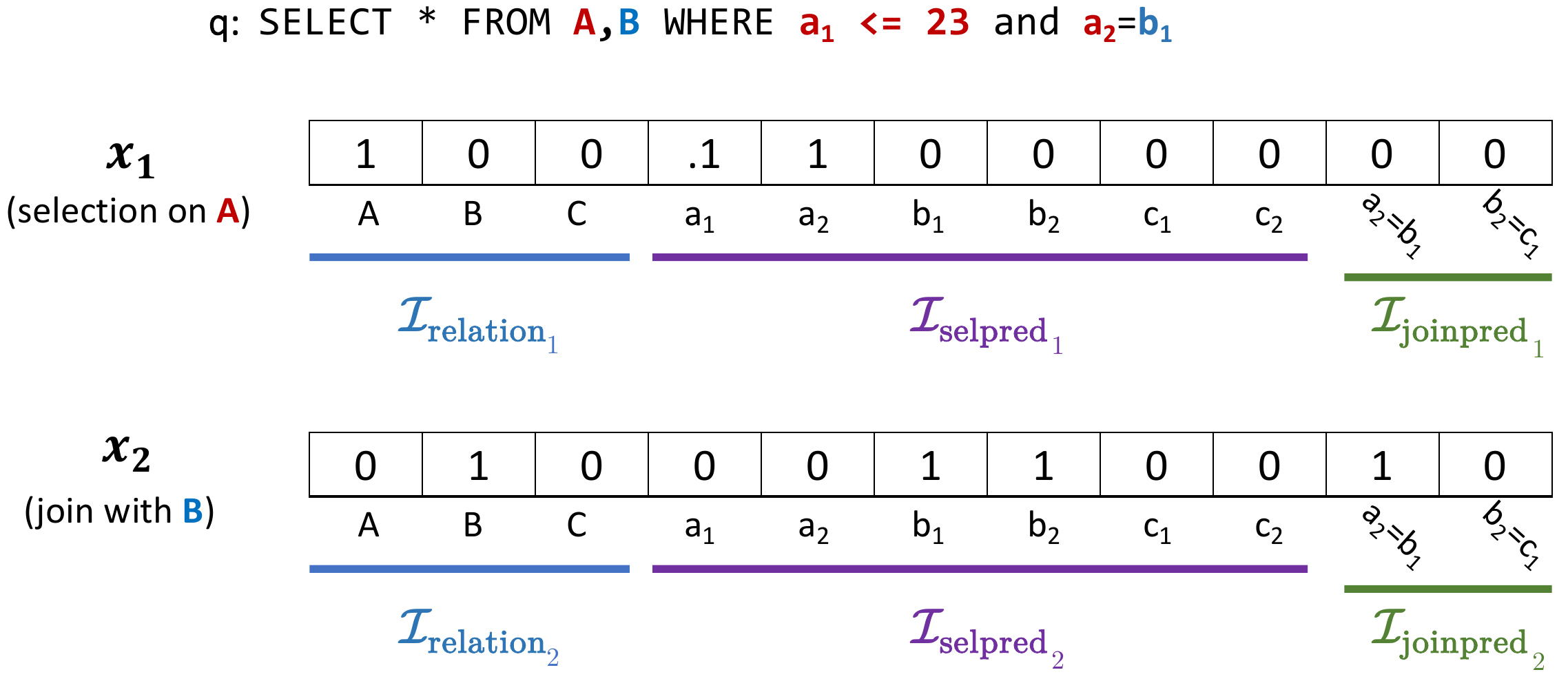}
    \caption{Query Encoding for the RNN Model: In this example, the input consists of two inputs: $\{x_1, x_2\}$. The first input represents the subquery that scans and filters relation A. The second represents the join with relation B.
    }
    \label{fig:rnn_join}
\end{figure}

To generate the input for this model, we concatenate three input vectors for \textit{each} action $x_i$. That is, for each action $x_i$, we concatenate vectors $\mathcal{I}_{relation_i}$, $\mathcal{I}_{selpred_i}$ and $\mathcal{I}_{joinpred_i}$. In \autoref{fig:rnn_join}, we show the representation for our running example, SELECT $*$ FROM $A, B$ WHERE $a_1 \leq 23$ and $a_2=b_1$. We break down this query into two operations: the scan and selection on relation $A$, followed by a join with relation $B$. Alternatively, we could have the first action represent the scan on relation $B$ (with no selections applied), followed by a join and selection with relation $A$.

\begin{figure}[t]
    \centering
    \includegraphics[width=.5\linewidth]{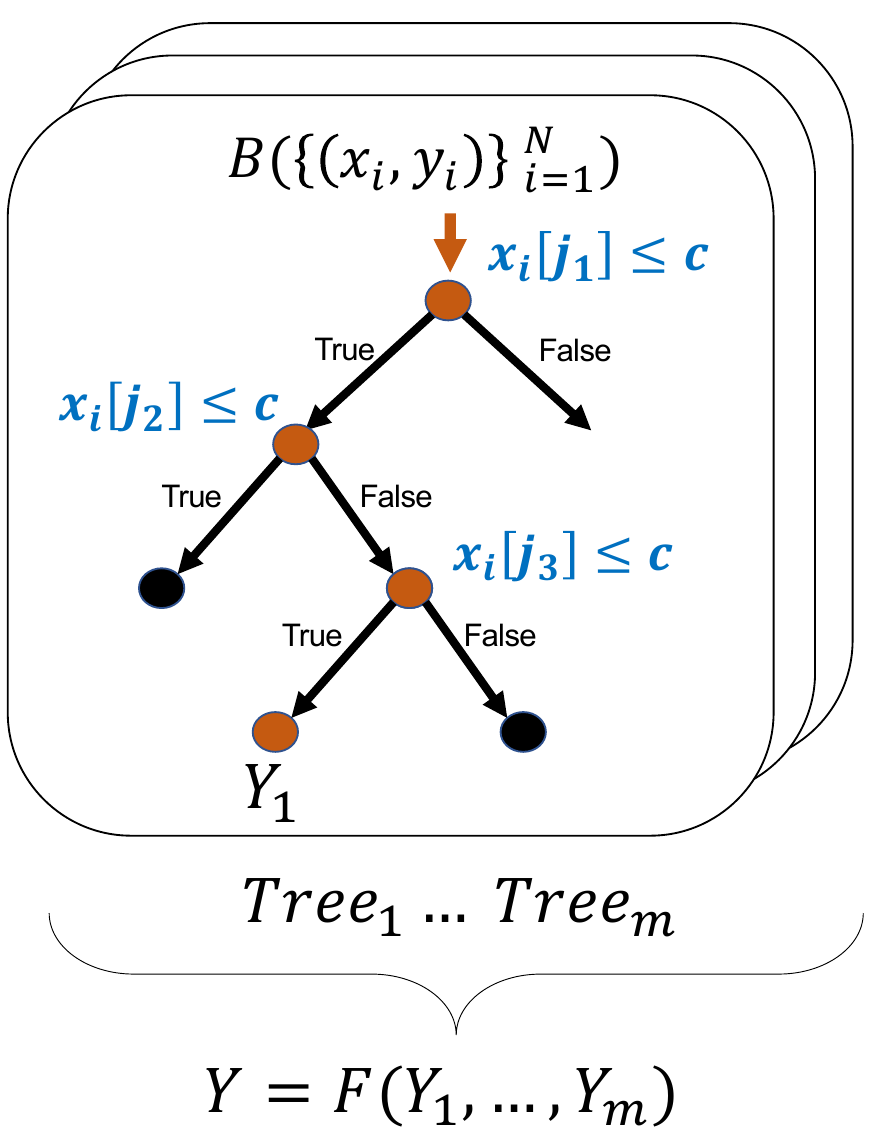}
    \caption{Illustration of a Random Forest Model: Before training, each tree is provided with a bootstrapped sample of training points, $B(\{(x_i, y_i)\}^N_{i=1})$. During inference, each $x_i$ is evaluated against the criteria in each node (where each $x_i[j]$ represents an attribute in $x_i$). The predictions from each tree are aggregated for the final prediction, $Y$.}
    \label{fig:trees}
\end{figure}

\begin{figure}[t]
    \centering
    \includegraphics[width=\linewidth]{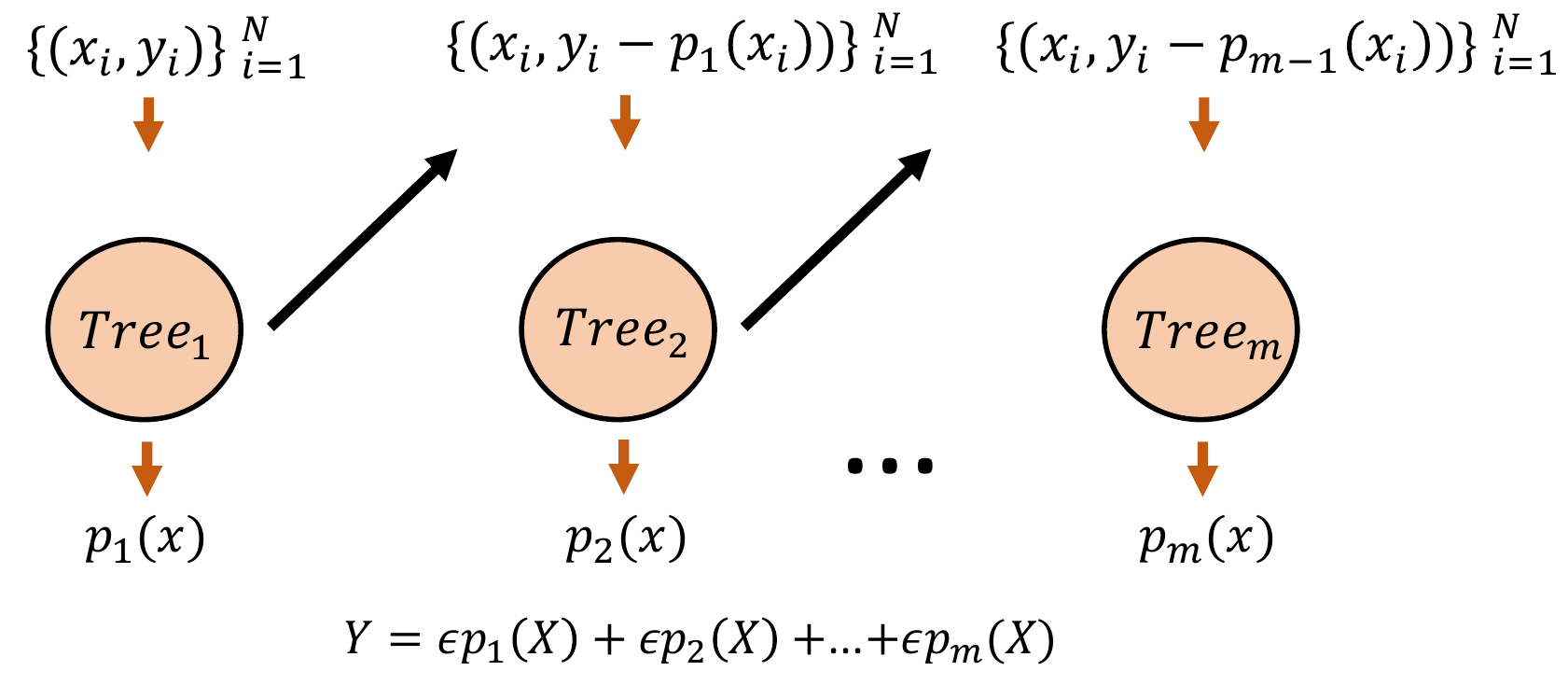}
    \caption{Illustration of a Gradient Boosted Tree Model: The first tree in the sequence is provided with $N$ data points from the training set. This tree learns the function $p_1(x)$. Subsequent trees are built incrementally based on the residual errors of the previous tree.}
    \vspace{-.5cm}
    \label{fig:boostedTrees}
\end{figure}

\subsection{Tree Ensembles}

In addition to deep learning models, we also include an analysis of tree ensembles. In particular, we compare \textit{bagging} and \textit{boosting} approaches. Ensemble methods aim to combine several decision models together to generate a single strong learner.

\textbf{Bagging} As a form of bagging, we study random forest models, as they are fast to build and are well suited for regression and classification problems~\cite{Criminisi:2012}. A random forest model is a combination of predictions from several independently trained trees as shown in \autoref{fig:trees}. During training, each tree takes as input a bootstrapped sample of $N$ training points, $B(\{(x_i, y_i)\}^N_{i=1})$ where $x_i$ and $y_i$ represent the input and label respectively. Each tree generated partitions the input space into subregions, where each of these subregions either contains a linear model~\cite{Quinlan:92} or a constant to make a prediction~\cite{Bishop:06}. Finding these subregions requires finding an optimal set of splits, which is computationally infeasible. Instead, these models use a greedy optimization to incrementally grow the trees one node at a time. To help generalize the model and to help add randomness, each tree generated uses a bootstrapped dataset from the training data~\cite{Denil:14}. Based on the predictions from all these trees, the random forest model uses a function $F$ (usually the mean) to compute a final prediction $Y$.

\textbf{Boosting} Another variant of ensemble models is known as \textit{boosting}, which involves training multiple models in a sequence. For AdaBoost, a well-known boosting algorithm, each training data point $i$ is initialized with a weight coefficient. These weight coefficients are adjusted according to the performance of the previously trained model~\cite{Bishop:06}. Data points with higher inaccuracies are given a greater weight, to emphasize fixing these errors in the next model. To implement a boosting technique in this work, we use gradient boosted trees. Instead of increasing the weights for each data point as seen in AdaBoost, the gradient boosted tree approach fits the next model based on the residual errors made by the current model~\cite{Parr:18}. We illustrate this approach in \autoref{fig:boostedTrees}. Initially, a single tree is built based on $N$ data points from the training set: $\{(x_i, y_i)\}^N_{i=1}$. Given the errors observed from this initial tree, the algorithm iteratively improves the errors by minimizing the residual errors in the next tree. The algorithm stops once it reaches a limit on the number of estimators or once the training objective does not change. As shown in the figure, to compute $Y$ given $X$, the predictions from all the trees are summed up (given a shrinkage parameter, $\epsilon$).

%% file: measurement_analysis.tex

\section{Measurement Analysis}~\label{sec:measurement_analysis}

In this section, we evaluate all models and their architecture variants on three datasets. We start with a description of the experimental setup, which includes the implementation of the models and generation of the training datasets. We then evaluate the accuracy, time, space trade-offs in \autoref{sec:joinselections_eval_datasets}, followed by a study of the robustness of the models in \autoref{sec:robustness} and a look into their latents in \autoref{sec:latents}.

\input{EvalFigure_Snippets/fig_cdf_storage_limit.tex}

\input{EvalFigure_Snippets/tab_percent_easy.tex}

\subsection{Experimental Setup}

\textbf{Datasets:} We evaluate the models on three datasets:

\begin{itemize}
    \item \textbf{IMDB}: The \textit{Internet Movie Data Base} is a real dataset that contains a wide variety of information about actors, movies, companies, etc. This dataset has 21 relations. The dataset is based on the 3.6GB snapshot from Leis et. al. \cite{Leis:17}. 
    \item \textbf{DMV}: This dataset contains 6 relations (61MB) and is based on a real-world dataset from a Department of Motor Vehicles~\cite{kiefer:17}. Relations include accidents, owners, cars, location, demographics and time.
    \item \textbf{TPC-H (skewed) }: This is a standard benchmark dataset with 8 relations and a scale factor of 1 (1GB). We adjust the skew factor to $z=1$~\cite{Chaudhuri:16}.
\end{itemize}

\textbf{Model Architectures:}
For the recurrent and neural networks, we build several models that vary in width (w) and depth (d). To minimize the number of possible architecture combinations, we assume that all layers within a model have the same width. We annotate the models as a pair ($x$, $y$), where $x$ represents the width and $y$ represents the depth. For example, (100w, 1d) represents a model with 100 hidden units in a single hidden layer. For the random forest models, we vary the the depth and the number of trees from 1 to 500. We similarly vary the number of trees, depth and shrinkage for the gradient boosted trees. We compare these models to estimates from PostgreSQL version 9.6~\cite{postgres}. To fairly compare PostgreSQL to these models in terms of space, we modify the PostgreSQL source to allow for a larger number of bins in each histogram. For each relation in each dataset, we collect statistics from each join predicate column and each selection column. We vary the number of bins from the default size (100 bins) up to 100K.

\textbf{Training Data:} For each dataset, we generate various training sets with different levels of query complexity. We define three complexity levels: $2Join$, $4Join$ and $6Join$. $2Join$ is the case where we generate a training set with joins that consist of any 2 relations in the dataset, $4Join$ represents joins with 4 relations, and $6Join$ represents joins with 6 relations. In addition, for each dataset, we manually select a set of columns as candidates for selection predicates. We select columns with small discrete domain sizes, as these are generally the columns that contain more semantically meaningful information about the data, unlike columns that contain a sequence of identifiers. As we generate the workload, selection predicate values are randomly drawn from the domains of the selected candidate columns. We generate 100K training samples for each query complexity and each dataset. We randomly select the desired number of tables and pick the selection columns from the joined relations.  For the RNN, because it requires an input for each timestep, we extend these training sets by adding more training samples for all the subqueries. For example, for a query that joins six relations, we extend the training set with additional examples representing the subquery after each intermediate join. For each query complexity training set, we select 1K samples to serve as the test set. 

\textbf{Hyperparameter Tuning:} We tune each model architecture for each dataset. We separate 10$\%$ of the training data as the validation data. We run a basic grid search over the learning rate and batch size. A larger batch size (although faster to train, especially on a GPU) might lead to sub-optimal results, while a small batch size is more susceptible to noise. Larger learning rates also have the tendency to oscillate around the optimum, while smaller learning rates might take a long time to train. We set the number of epochs to 500 for all learning rate and batch parameter combinations. Based on the combination that leads to the lowest learning rate, we continue to train for more epochs as long as the validation loss keeps decreasing. We stop the training once the validation loss plateaus or increases.

\textbf{Model Implementation Details:}
The neural network is implemented in Tensorflow~\cite{Abadi:16} and is implemented as a residual network  with leaky RELU activation functions, as it is a default recommendation to use in modern neural networks~\cite{Goodfellow:16}. Weights are initialized from a random normal distribution with a small standard deviation. Biases are initialized to .01. The input $X$ is normalized as explained in \autoref{sec:nn} and centered using a StandardScaler. The output $Y$ is log transformed and also normalized with a StandardScaler. The model's goal is to minimize the mean squared error between the real outputs and the predictions. We use the AdamOptimizer as the optimizer for the model. The recurrent neural network is also implemented in Tensorflow. For deep recurrent neural networks, we use a ResidualWrapper around each layer, to mimic the residual implementation of the neural networks. Both the neural network and recurrent neural networks are run on a GPU on p2.xlarge instances on Amazon AWS~\cite{amazonaws}. The Random Forest model is based on an implementation from sklearn's RandomForestRegressor module~\cite{scikit-learn}. Finally, the gradient boosted tree model is based on sklearn's GradientBoostingRegressor~\cite{scikit-learn}.

\subsection{Learning Cardinalities for Selections + Joins}\label{sec:joinselections_eval_datasets}

In this section, we vary the architecture of the models and evaluate them on the three datasets. We study the trade-offs (space, time, and accuracy) for these models.

First, we evaluate the prediction accuracy for each model. As described in \autoref{sec:background}, we make the assumption that the query workload is known in advance (we relax this assumption later in this section). In this case, the models overfit to a specific set of queries. As a result, training overfit models helps us study how effectively these models are able to compress information. For each query complexity, we train six neural network (NN) and six recurrent neural network models (RNN) based on the following widths and depths: (100w, 1d), (100w, 5d), (500w, 1d), (500w, 5d), (1000w, 1d), (1000w, 5d). We separately train random forest models and gradient boosted tree models with 1, 5, 50 and 500 trees. Larger models generally use up more space, but result in more accurate cardinality predictions. 

To make this analysis comparable to PostgreSQL, we first limit the storage budget for the models to be no more than the storage budget for PostgreSQL histograms. We compute the size of a model as the size of all its parameters. For the NN and RNN models, we thus measure the number of trainable variables and for PostgreSQL, we measure the number of parameters used in the $pg\_stats$ table. We compare PostgreSQL cardinality estimates to those produced by models that are smaller in size compared to PostgreSQL's histograms. We specifically study the PostgreSQL scenario where each histograms builds at most 1K bins. Setting the number of bins to 1K for PostgreSQL results in 13385 parameters for the DMV dataset, 15182 parameters for the TPC-H dataset and 44728 parameters for the IMDB dataset. We purposely set PostgreSQL as the storage upper bound size. Given these storage budgets, we then select the best neural network architecture, the best recurrent network architecture, the best random forest model and best gradient boosted tree model. If no model meets the budget, we do not display them on the graphs. If more than one model architecture meets the storage budget, we display the best model, where the best model is defined as the one with the lowest median error.

\textbf{Limited Storage CDFs and Outlier Analysis:}
For PostgreSQL, as for other relational DBMSs, cardinality estimation is easy for some queries and hard for others. As expected PostgreSQL yields more accurate predictions for queries with a low complexity, particularly those with no selection predicates. To help distinguish between these ``easy'' and ``hard'' queries (labeled as Easy(PostgreSQL) and Hard(PostgreSQL)), we plot the absolute errors from PostgreSQL as a cumulative distribution (cdf) as shown in \autoref{fig:cdf_knee}. We use the knee ($k$) of the cdf curve to split the queries into an  ``easy'' category (those with errors less than the knee, $k$) and a ``hard'' category (those with errors greater than the knee $k$). For the TPC-H dataset, the distribution of errors is wide. To ensure we retain enough queries in the Hard(PostgreSQL) category (for later more in-depth analysis), we compute $k$ and half the corresponding error. 

\input{EvalFigure_Snippets/tab_hard_2join.tex}

\input{EvalFigure_Snippets/fig_column_distributions.tex}

\input{EvalFigure_Snippets/fig_cdf_unlimited_storage.tex}

\input{EvalFigure_Snippets/fig_trade-offs.tex}

We first focus on the Hard(PostgreSQL) queries. These are the more interesting queries to study as these are the queries for which we seek to improve cardinality estimates. We plot the distribution of errors for the best performing models for each dataset in \autoref{fig:cdf_storage_limit}. Overall, both types of models outperform PostgreSQL on all three datasets. We also find the performance of both types of models to be similar.

First, in \autoref{fig:storage_imdb}, we show the cdf for the Hard(PostgreSQL) queries from the IMDB dataset. From the entire set of IMDB queries, 11\% of the queries fall in the Hard(PostgreSQL) category. The y-axis represents the percentage of queries and the x-axis represents the absolute error. In addition to the PostgreSQL error curve, we show the cdf for the corresponding queries from the best neural network and recurrent neural network models. We do not show the tree ensemble models here, as the models do not meet the storage budget. Both the neural network and recurrent neural network have comparable cardinality estimation errors. On average, the neural network reduces estimation error by 72\%, while the recurrent neural network reduces the error by 80\%. Below \autoref{fig:storage_imdb}, we include additional details that show the percentiles of the model cdfs, the average absolute error for each query complexity, and the average relative error. 

In \autoref{fig:storage_dmv}, we show the cdf for Hard(PostgreSQL) from the DMV dataset. Approximately 10\% of the queries are labeled as hard for PostgreSQL. The NN reduces the errors by 75\% on average and the RNN by 73\%. As shown in the tables below the figure, the complexity of the queries does not heavily impact the average error. In fact, the relative errors for the NN across all query complexities have a small standard deviation ($\sigma=.004$), compared to IMDB ($\sigma=.08$). Compared to DMV, the IMDB dataset contains several many-to-many primary/foreign key relationships, so joining relations significantly increases the size of the final join result. 


\input{EvalFigure_Snippets/fig_robust_selections.tex}

We also observe a significant error reduction in \autoref{fig:storage_tpch} (TPC-H), where the NN improves estimates by 98\% and the RNN by 97\%. For TPC-H, 30\% of the queries are hard for PostgreSQL. 

\autoref{tab:percent_easy} shows the percentage of queries that are easy for the models given that they are either Easy(PostgreSQL)  or Hard(PostgreSQL) for PostgreSQL. In the case of Hard(PostgreSQL) queries, 70\% or more become easy with the models. For the Easy(PostgreSQL) queries, the simple NN and RNN models also find a majority of these queries to be easy (>90\%). For IMDB and DMV, there are some queries from the Easy(PostgreSQL) batch that the models find to be hard. We highlight some of these hard queries below:

\begin{itemize}
    \item From the IMDB dataset, approximately 0.4\% of the Easy(PostgreSQL) queries are hard for the NN, and we find that the query with the highest error is one with an absolute error of 8.9M. This query joins the $name$, $cast\_info$, $role\_type$, and $char\_name$ relations and has a selection predicate on $role\_id <= 8$. For the RNN, the query with the highest error is similar. It joins the $name$, $cast\_info$, $role\_type$, and $char\_name$ relations, with a selection predicate on $role\_id<=4$.
    \item For the DMV dataset, approximately 10.5\% of the queries are hard for NN, and 6.5\% are hard for the RNN. The query with the highest error for the NN is one that joins all relations $car$, $demographics$, $location$, $time$, $owner$, and $accidents$ and has several selection predicates: $age\_demographics<=89$, $month\_time<=12$, $year\_accidents<=2004$. For the RNN, the query with the highest absolute error also joins all relations and has selection predicates with similar values, $age\_demographics<=93$,$month\_time<=9$,$year\_accidents<=2005$.
\end{itemize}

From the Hard(PostgreSQL) queries, there are more queries that remain difficult for the models compared to Easy(PostgreSQL). These hard queries consist of joins of 6 relations (the most complex queries we have in the test set) and up to 5 selection predicates. 

Understanding why the NN or RNN fail to accurately predict the cardinality for specific queries is challenging as there are several factors to consider. For example, the error could be caused by a specific join or perhaps a combination of selection attributes. To gain a better understanding of these errors, we now only focus on the queries with a low complexity (i.e. those from the $2Join$ test set). In \autoref{tab:hard_2join}, we take the Easy(PostgreSQL) queries and show the queries with the highest errors from the $2Join$ set. For succinctness, we annotate each query with the names of the relations it joins (relations are listed in parenthesis) and its selection predicates. We further add the absolute error of the query in brackets.

For IMDB, the hardest queries for the NN and RNN are similar. These queries consist of joins with $cast\_info$ and either $role\_type$ or $name$. All queries also have a selection predicate on the $role\_id$ column with values between 6 and 11.  \autoref{fig:column_distributions} shows the value distributions for different attributes. The first row shows all selection columns for IMDB, the second for DMV, and third for TPC-H. The x-axis in each graph represents the column value and the y-axis represents the frequency of the value. In \autoref{fig:roleid}, we show the distribution of the $role\_id$ column. The red bars represent the values for which we see the highest errors for the NN and RNN models, based on \autoref{tab:hard_2join}. Compared to the other existing selection attributes, $role\_id$ comes from the largest relation in the dataset, $cast\_info$. We generally observe that the models have the highest errors for columns that belong to the largest relations and specifically at the points where the distribution is irregular. 

For the DMV dataset, the hardest queries are those that contain the $accidents$ relation and join with $time$ or $location$. These queries have selection predicates on the $year$ and $month$ columns. We highlight the selection predicate values in \autoref{fig:year} and \autoref{fig:month}. We note that there is a one-to-one mapping between the $accidents$ and $time$ relation, so the distribution for these columns does not change due to the join. This is also the case for the join between $accidents$ and $location$. The year column in the accidents relation has a high skew and the models have the highest errors for the more frequent values. The accidents relation also happens to be the largest relation in the DMV dataset.


For the TPC-H dataset, most of the errors come from the join between $lineitem$ and $orders$. These contain selection predicates on both the $l\_linenumber$ and $l\_quantity$. The pearson correlation for these two attributes is low ($.0002$) so these are independent attributes. We highlight the values with highest errors in \autoref{fig:lquantity} and \autoref{fig:linenumber}. We note that the $l\_quantity$ in particular has an irregular distribution, and also belongs to the relation with the highest number of tuples in the dataset, $lineitem$.


\textbf{Unlimited Storage CDFs and Outlier Analysis} In \autoref{fig:cdf_unlimited_storage}, we show similar graphs across all datasets, but with an unlimited storage budget. The goal here is to understand how more complex models compare against the simpler ones from \autoref{fig:cdf_storage_limit}. Given this unlimited budget, we now include the tree ensemble models. The PostgreSQL estimates do not significantly change even with 100K bins, which implies that adding finer granularity to the histograms does not significantly improve estimates. Among all the models, the trees (in particular, the gradient boosted trees) have the lowest errors overall across all query complexities and across all datasets.


\textbf{Time vs Space vs Accuracy Trade-offs} In \autoref{fig:trade-offs}, we show the error, space and time trade-offs for each model. First, in \autoref{fig:space} we compare the error and space. On the y-axis, we show the absolute error between the predicted value and the real value on a log scale. On the x-axis we show the space of each model on a log scale. Each point represents the median error and the error bars represent the 25th and 75th percentiles. For all datasets, all variants of PostgreSQL have the highest errors and increasing histogram bin granularity does not significantly improve performance. Neural networks and recurrent neural networks are fairly competitive in terms of absolute error. In \autoref{sec:practical}, we study whether deeper models actually learn more context about the relations compared to the shallower ones.  Models that are deeper are much larger in terms of space, with small error improvement over simpler models. 

In \autoref{fig:time}, we compare the accuracy to the time (in seconds) it takes to train each model. We do not include the time it takes to run the hyperparameter tuning and we do not include the time it takes to run the training queries. 


Given their large sizes, an important question is whether the models improve upon simply keeping the entire query workload in a hash table (with query features as keys and cardinalities as values). To answer this question we plot the overhead of such a hash table. Given that our training data consists of only 100K samples for each query complexity, our goal is to understand whether the deep learning models can actually compress information and still provide a good accuracy. For the hash table model, we assume that each feature for each training example is equivalent to one weight when measuring space. To measure time, we measure the time it takes to populate the hash table. We mark this implementation in the graphs as a vertical dashed line. For this model, the error is 0 for each query. 

For each dataset, all variants of the tree ensemble models result in the lowest errors.  For the random forest model, trees with the lowest error are those with a single decision tree. Since we build these models to overfit to a specific query workload, using a single decision tree results in the lowest error. Once more decision trees are introduced, the error is higher as these models no longer overfit and attempt to generalize over the training set. For the boosted trees models, generating more trees incrementally lowers the residual errors, and as a result, does not impact the overfitting. This also  depends on how we tune the boosted tree models. Since our goal is to overfit, the best boosted tree models are those that contain a high depth and high shrinkage rate. These results suggest that for overfit workloads, the ensemble models are able to build these models quickly  and more accurately compared to the deep learning models. The deep learning models are able to save in space and although they are not as accurate as the trees, they can still improve errors in some cases by an order of magnitude compared to PostgreSQL.

\input{EvalFigure_Snippets/fig_robust_joins.tex}

\subsection{Model Robustness} \label{sec:robustness}

In this section, we study how robust these models are in the face of \textit{unknown} queries. That is, instead of overfitting each model to a specific set of queries, we remove some query samples from the training data. We focus on the most challenging, the $6Join$ queries. We evaluate the most complex models for the RNN and NN (1000w, 5d) as these perform favorably for the $6Join$ set. We also select the best performing version of the random forest model and booted tree models.

\textbf{Removing Selections} In the first row of \autoref{fig:robust_selections}, we remove 10\% of values from three columns (one from each dataset): $production\_year$ (IMDB) in \autoref{fig:im_remove_selection_1}, $year$ (DMV) in \autoref{fig:d_remove_selection_1} and $l\_quantity$ (TPC-H) in \autoref{fig:t_remove_selection_1}. As shown in \autoref{fig:im_remove_selection_1}, for the IMDB dataset, the accuracy of the tree ensembles outperforms the other models. For both the DMV and TPC-H dataset (shown in columns 2 and 3) of \autoref{fig:robust_selections}, the NN and RNN models turn out to be more robust compared to the trees. In these graphs, we also included the accuracy of the hash table model implementation. Since the data points in the test set are not included in the training set, we use a nearest neighbor approach to find the closest sample that exists in the training set (stored in the hash table). We use the nearest neighbor implementation from sklearn~\cite{scikit-learn} which uses the minkowski distance metric. In many cases, the hash model performs similarly to the tree ensemble models, except for IMDB, where the hash model is not as accurate. For the IMDB dataset, we generate 100K random query samples uniformly from the set of all possible queries, but unlike the other two datasets, 100K queries doesn't fully cover the set of all possible queries for this dataset. As a result, the nearest neighbor is not always as close for this database as it is for the other two.


\textbf{Removing Joins} In \autoref{fig:im_remove_join} and \autoref{fig:t_remove_join}, we remove a join with a specific combination of tables from the IMDB and TPC-H $6Join$ datasets. During training, the models observe how certain tables join with each other, but they never see the specific combinations we remove. In \autoref{fig:im_remove_join}, we remove the join between relations: $\{$ $complete\_cast$, $aka\_title$, $movie\_info\_idx$, $title$, $movie\_companies$, $movie\_link$ $\}$ from the training set. The queries shown in the figure correspond to the test set, which includes the removed combination of tables with random selection predicates. For the IMDB dataset, we observe that the tree models rely heavily on features from $\mathcal{I}_{selpred}$. We found that the IMDB dataset contains combinations of tables in the training data that are very similar (and yield the same cardinality) as the combination of tables we removed from the training. The hash table model has the worst accuracy, since the nearest neighbor at times selects queries with selection predicates on the same values but different underlying tables.
 



In \autoref{fig:t_remove_join}, we observe a similar trend. For this experiment, we remove a join from the TPC-H dataset with the relations: $\{$ $customer$, $lineitem$, $partsupp$, $nation$, $part$, $orders$ $\}$. For this dataset, the RNN and NN models are more accurate compared to the tree ensembles and the hash table model. 

Ultimately, these graphs show that the deep learning models are generally more robust in scenarios where we are not simply overfitting the models to known training data.


\subsection{Model Latents}~\label{sec:latents}
One challenge of training deep neural networks is the difficulty to understand what the models are actually learning. As discussed in \autoref{sec:robustness}, the tree models are easily interpretable as we can track path of decision splits to understand how the model is able to predict the outcome given the input. For neural networks, diagnosing why a model arrives at a specific answer is a harder problem. There are several existing approaches, which include masking or altering the input to measure the predication change and studying hidden unit activation values~\cite{Kahng:18,Selvaraju:17,Zeiler:14}.

We study the activation values of the hidden layers for the NN and RNN models. During training, these models take the input, $X$, and propagate it through a series of transformations that represent the data at different levels of abstraction~\cite{Goodfellow:16}. Taking a close look at the activation values (also referred to as \textit{latent representations} or \textit{embeddings}) can help diagnose what the model is learning from the inputs. For example, if we cluster training samples based on their latents, we can determine whether models are in fact generating similar representations for queries that are semantically similar. 

We use the t-SNE technique to cluster latents, which is a dimensionality reduction technique that is often used to visualize high-dimensional data~\cite{Maaten:08}. This approach has an objective function that minimizes the KL-divergence between a distribution that measures pairwise similarities of the objects in high-dimensional space and a distribution that measures the pairwise similarities of the corresponding low-dimensional points~\cite{Maaten:08}. Compared to principal component analysis (PCA), t-SNE is non-linear and relies on probabilities. 

We cluster latent vectors from the (100w, 1d) NN model for the $6Join$ training set from each dataset. In \autoref{fig:latent_representations}, we reduce the dimensionality of the latents from the (100w, 1d) model on the TPC-H dataset (100 hidden units total) down to three dimensions, which is the highest number of dimensions allowed for t-SNE. In the figure, there are four clusters, each representing different sets of joins:

\begin{itemize}
    \item Cluster 1: customer,lineitem,nation,orders,partsupp,region 
    \item Cluster 2: customer,lineitem,orders,part,partsupp,supplier
    \item Cluster 3: customer,lineitem,nation,orders,partsupp,supplier
    \item Cluster 4: customer,lineitem,nation,orders,part,partsupp 
\end{itemize}

For t-SNE, the distance between clusters is irrelevant, the more important factor is the relevance among the points that are clustered together. For the DMV dataset and IMDB, the clusters do not represent combinations of relations, but we observe that queries that are near each other share similar selection predicate values.

\input{EvalFigure_Snippets/fig_latent_representations.tex}

\input{EvalFigure_Snippets/fig_speed_up.tex}

For the RNN (100w, 1d) model, we find that clusters are determined based on the sequence of operations. Recall, during training, the RNN learns to predict cardinalities for different join sequences, as a result of observing many queries. We observe that the resulting clusters represent queries that end with similar operations. For example, one cluster contains combinations of relations $orders$, $lineitem$, $partsupp$, but always ends the sequence with joins on either the $supplier$ and $part$ relation or $customer$ and $supplier$. We find that complex models (1000w, 5), also show a similar trend. This is actually a side-effect of RNNs, as more recent actions have a heavier influence on the content that exists in the hidden states. More specifically, it is difficult to learn long-term dependencies as the gradient is much smaller compared to short-term interactions~\cite{Goodfellow:16}.

As an additional experiment, we cluster the latents from queries that have not been included in the training. Ideally, although these queries have never been observed by the model, they should cluster with similar training queries. We focus on the TPC-H join removal scenario, originally shown in \autoref{fig:t_remove_join}. When we cluster the latents from the (1000w, 5d) NN model, the queries that were not included in the training are clustered separately from the rest. This seems to imply that the NN does not learn the interactions between subqueries. This is not the case for the RNN, as  queries that are left out of training are clustered together with queries that have similar subqueries. For example, a query that joins relations $lineitem$, $orders$, $partsupp$, $customer$, $part$, and $nation$, is clustered together with queries that contain relations $lineitem$, $orders$, $partsupp$, $customer$, $part$, and $supplier$.

%% file: EvalFigure_Snippets/fig_cdf_storage_limit.tex
\begin{figure*}[!htbp]
\centering 
\begin{subfigure}{.30\linewidth}
\includegraphics[width=\linewidth]{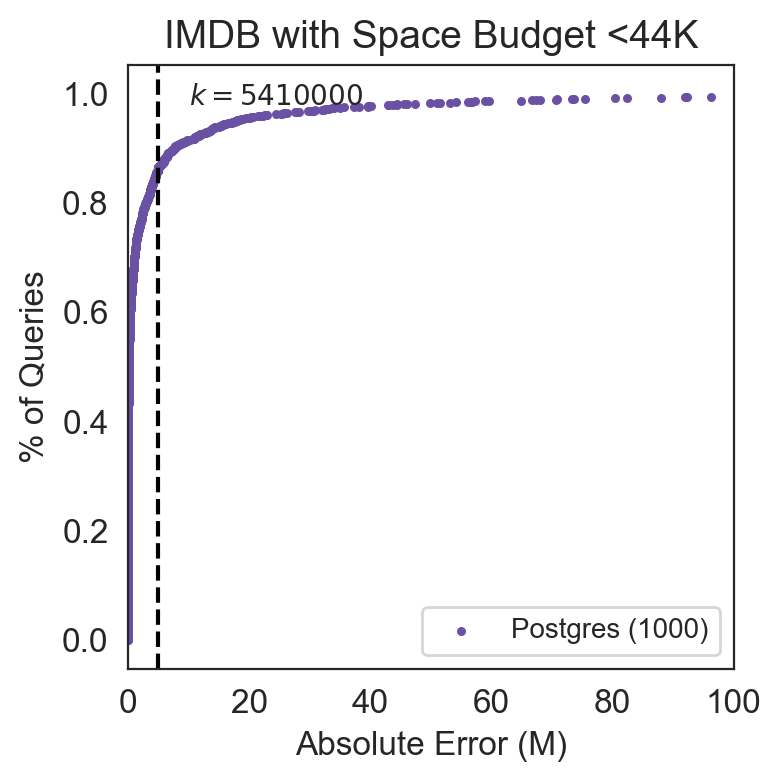}%
\label{fig:storage_imdb}
\end{subfigure}
\begin{subfigure}{.30\linewidth}
\includegraphics[width=\linewidth]{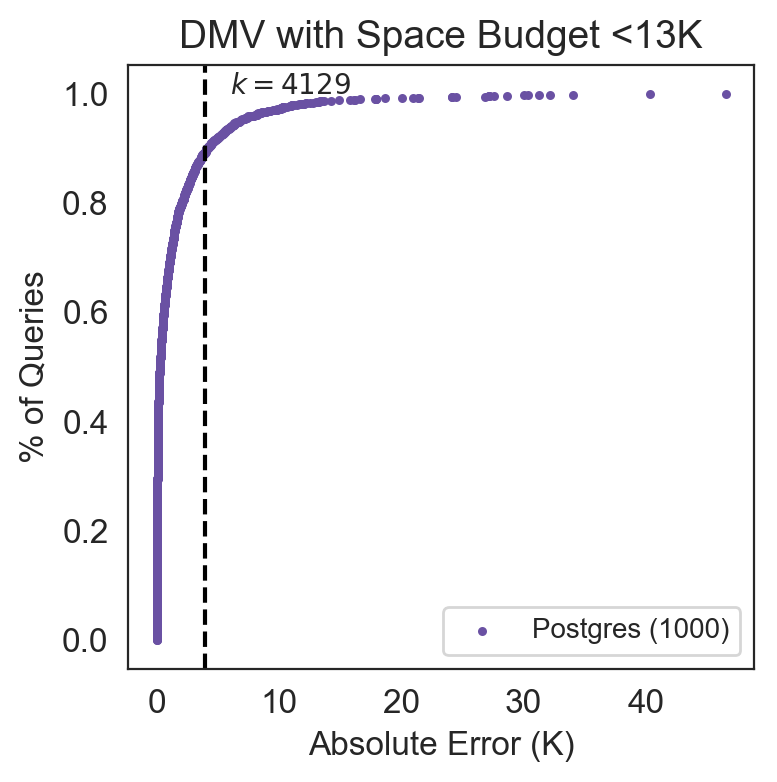}%
\label{fig:storage_dmv}
\centering
\end{subfigure}
\begin{subfigure}{.30\linewidth}
\includegraphics[width=\linewidth]{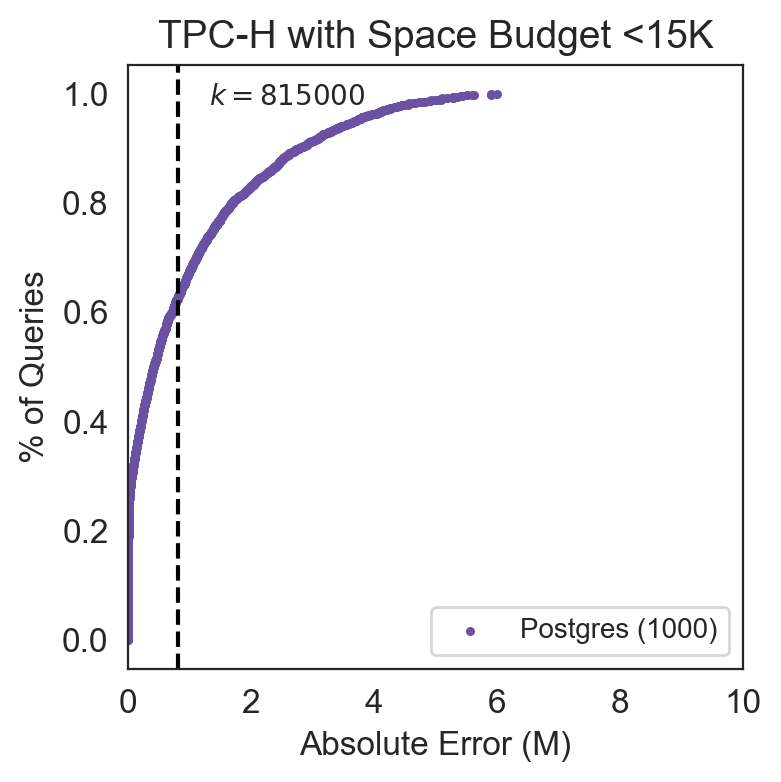}%
\label{fig:storage_tpch}
\end{subfigure}\\
\caption{CDF of PostgreSQL absolute errors with storage budget: For each curve, we show the knee, $k$, which defines the split between Easy(PostgreSQL) and Hard(PostgreSQL).}
\label{fig:cdf_knee}

\bigskip
\bigskip
\bigskip

\centering
\begin{subfigure}{.30\linewidth}
\includegraphics[width=\linewidth]{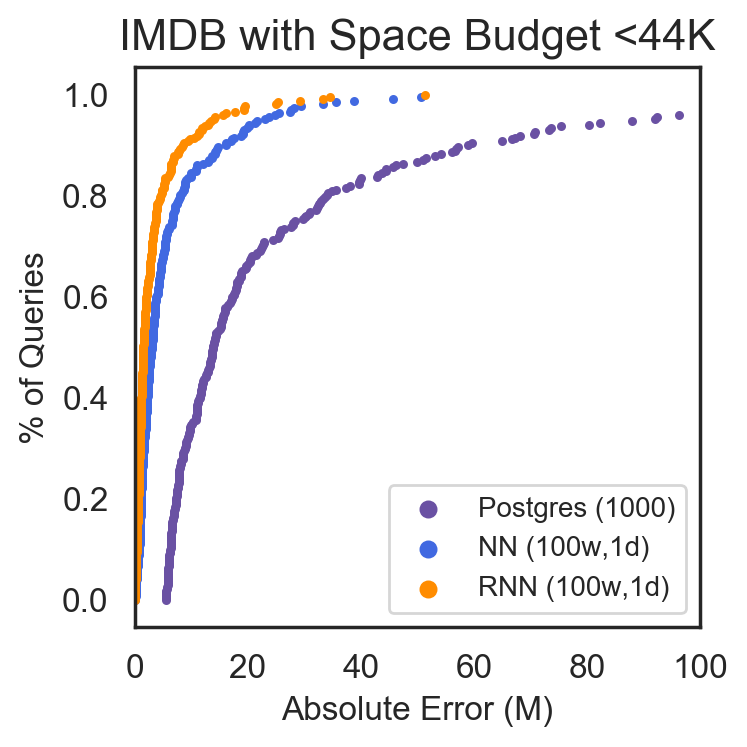}%
\caption{IMDB with Storage Budget}%
\label{fig:storage_imdb}

\centering
\scalebox{0.8}{
\begin{tabular}{|l|lll|}
\hline
\multicolumn{4}{|c|}{CDF Percentiles}       \\ \hline
           & $25\%$ & $50\%$ & $75\%$ \\ \hline
PostgreSQL & 7.8M & 13.8M & 29.7M \\ 
NN (100w,1d)  & 1.25M  & 2.97M & 6.6M    \\
RNN (100w,1d) & .71M &  1.49M & 3.67M   \\ \hline
\end{tabular}
}

\bigskip

\centering
\scalebox{0.8}{
\begin{tabular}{|l|lll|}
\hline
\multicolumn{4}{|c|}{Average Absolute Errors}       \\ \hline
         & $2Join$ & $4Join$ & $6Join$ \\ \hline
PostgreSQL & 6.8M & 12.8M & 31.3M \\ 
NN (100w,1d)  & .80M   & 4.1M & 7.0M   \\
RNN (100w,1d) & .58M & 2.2M & 4.1M   \\ \hline
\end{tabular}
}

\bigskip

\scalebox{0.8}{
\begin{tabular}{|l|lll|}
\hline
\multicolumn{4}{|c|}{Average Relative Errors}       \\ \hline
           & $2Join$ & $4Join$ & $6Join$ \\ \hline
PostgreSQL & .39 & .75 & .95 \\ 
NN (100w,1d)  & .04  & .22 & .20    \\
RNN (100w,1d) & .03 & .11 & .13   \\ \hline
\end{tabular}
}

\end{subfigure}
\begin{subfigure}{.30\linewidth}
\includegraphics[width=\linewidth]{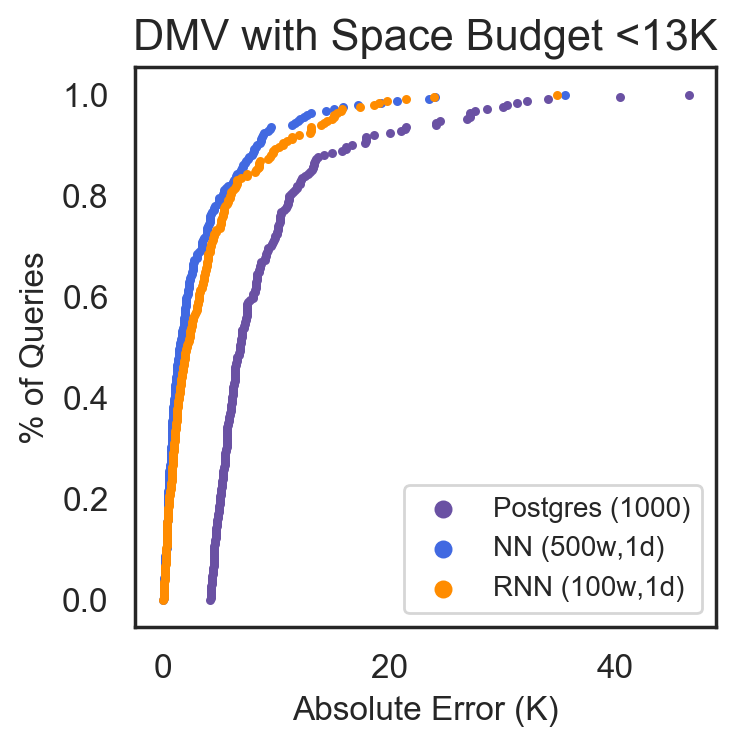}%
\caption{DMV with Storage Budget}%
\label{fig:storage_dmv}

\centering
\scalebox{0.8}{
\begin{tabular}{|l|lll|}
\hline
\multicolumn{4}{|c|}{CDF Percentiles}       \\ \hline
           & $25\%$ & $50\%$ & $75\%$ \\ \hline
PostgreSQL & 5.2K & 6.8K & 10.3K \\ 
NN (500w,1d)  & .50K & 1.4K & 4.1K   \\
RNN (100w,1d) & .76K & 2.1K & 5.1K   \\ \hline
\end{tabular}
}

\bigskip

\centering
\scalebox{0.8}{
\begin{tabular}{|l|lll|}
\hline
\multicolumn{4}{|c|}{Average Absolute Errors}       \\ \hline
         & $2Join$ & $4Join$ & $6Join$ \\ \hline
PostgreSQL & 9.4K & 9.4K & 8.9K \\ 
NN (100w,1d)  & 4.9K & 3.0K & 3.1K  \\
RNN (100w,1d) & 7.9K & 2.9K & 4.0K   \\ \hline
\end{tabular}
}

\bigskip

\scalebox{0.8}{
\begin{tabular}{|l|lll|}
\hline
\multicolumn{4}{|c|}{Average Relative Errors}       \\ \hline
           & $2Join$ & $4Join$ & $6Join$ \\ \hline
PostgreSQL & .10 & .20 & .23 \\ 
NN (100w,1d)  & .03  &  .03 & .04    \\
RNN (100w,1d) & .06  &  .02 & .07  \\ \hline
\end{tabular}
}

\end{subfigure}
\begin{subfigure}{.30\linewidth}
\includegraphics[width=\linewidth]{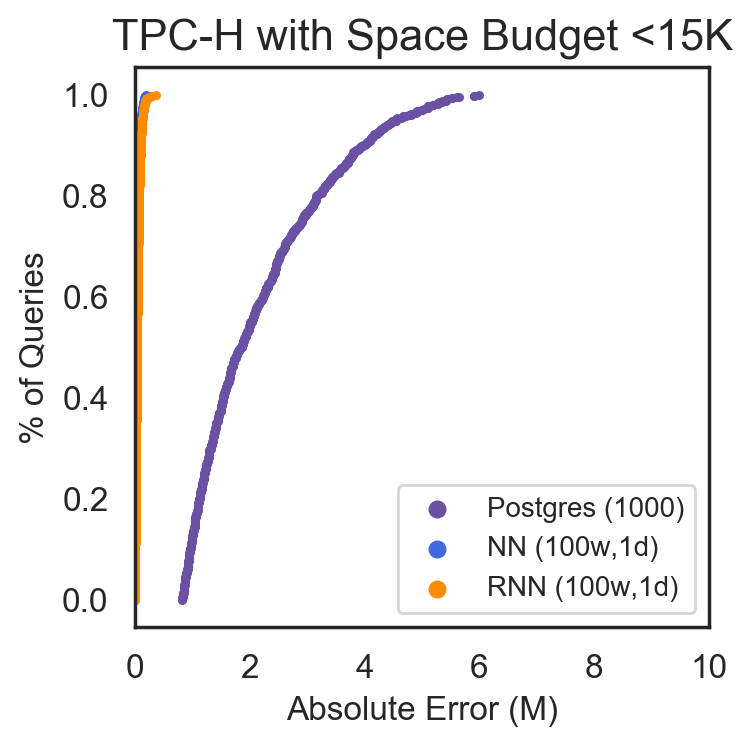}%
\caption{TPCH with Storage Budget}%
\label{fig:storage_tpch}
\centering
\scalebox{0.8}{
\begin{tabular}{|l|lll|}
\hline
\multicolumn{4}{|c|}{CDF Percentiles}       \\ \hline
           & $25\%$ & $50\%$ & $75\%$ \\ \hline
PostgreSQL & 1.2M & 1.8M & 2.9M\\ 
NN (100w,1d)  & .01M  & .02M & .05M    \\
RNN (100w,1d) & .01M & .03M  & .06M    \\ \hline
\end{tabular}
}

\bigskip

\centering
\scalebox{0.8}{
\begin{tabular}{|l|lll|}
\hline
\multicolumn{4}{|c|}{Average Absolute Errors}       \\ \hline
         & $2Join$ & $4Join$ & $6Join$ \\ \hline
PostgreSQL & 2.9M & 2.2M & 1.8M\\ 
NN (100w,1d)  & 35K & 40K & 32K\\
RNN (100w,1d) & 27K & 60K & 41K   \\ \hline
\end{tabular}
}

\bigskip

\scalebox{0.8}{
\begin{tabular}{|l|lll|}
\hline
\multicolumn{4}{|c|}{Average Relative Errors}       \\ \hline
           & $2Join$ & $4Join$ & $6Join$ \\ \hline
PostgreSQL & .99 & .99 & .99 \\ 
NN (100w,1d)  & .01  & .02 & .01    \\
RNN (100w,1d) & .01 &  .03 & .02  \\ \hline
\end{tabular}
}
\end{subfigure}\\

\caption{Error Analysis for all Models : We show the curve for Hard(PostgreSQL) and show the corresponding errors from the best models below the storage budget. Below each graph, we show tables detailing the percentiles, the average absolute error and average relative error.}
\label{fig:cdf_storage_limit}%

\end{figure*}

%% file: EvalFigure_Snippets/tab_percent_easy.tex
\begin{table*}[t]
\begin{tabular}{l|l|l|l|l|l|l|}
\cline{2-7}
                                              & \multicolumn{6}{c|}{\textbf{\begin{tabular}[c]{@{}c@{}}\% Queries Easy\\  (Models)\end{tabular}}} \\ \cline{2-7} 
                                              & \multicolumn{2}{c|}{IMDB}        & \multicolumn{2}{c|}{DMV}       & \multicolumn{2}{c|}{TPC-H}    \\ \cline{2-7} 
                                              & NN(100w,1d)    & RNN(100w,1d)    & NN(500w,1d)   & RNN(100w,1d)   & NN(100w,1d)   & RNN(100w,1d   \\ \hline
\multicolumn{1}{|l|}{\textbf{Easy(PostgreSQL)}} & 99.5\%         & 99.8\%          & 90.5\%        & 94.5\%         & 100\%         & 100\%         \\ \hline
\multicolumn{1}{|l|}{\textbf{Hard(PostgreSQL)}} & 71.4\%         & 83.5\%          & 75.6\%        & 69.4\%         & 100\%         & 100\%         \\ \hline
\end{tabular}
\caption{\textbf{Percentage of Queries that are Easy for the Models: For each Easy(PostgreSQL) query batch, we find the percentage of queries that are also easy for the models. We also show the percentage of queries that are easy based on the Hard(PostgreSQL) batch}}
\label{tab:percent_easy}
\end{table*}

%% file: EvalFigure_Snippets/tab_hard_2join.tex

\begin{table*}[t]
\scalebox{0.8}{
\begin{tabular}{c|l|l|}
\cline{2-3}
\multicolumn{1}{l|}{}       & \multicolumn{2}{c|}{\textbf{Queries with Highest Errors from $2Join$}}                                                                                                                                                                                                                                                                                                                                                                                                                                                                                                                                                                                                      \\ \cline{2-3} 
\multicolumn{1}{l|}{}       & \textbf{Best NN per Dataset}                                                                                                                                                                                                                                                                                                & \textbf{Best RNN per Dataset}                                                                                                                                                                                                                                                                                                                 \\ \hline
\multicolumn{1}{|c|}{IMDB}  & \begin{tabular}[c]{@{}l@{}}

(cast\_info,role\_type) where role\_id\textless{}= 11 {[}1.9M{]}\\ (cast\_info,role\_type) where role\_id \textless{}= 10 {[}1.9M{]}\\ (cast\_info,role\_type) where role\_id \textless{}= 8 {[}1.9M{]} \\
(cast\_info,title) where kind\_id <= 1,production\_year <=\textbackslash \\ 2019,role\_id<= 4 {[}1.5M{]} \\
(movie\_info,info\_type) {[}1.3M{]} \\
(cast\_info,role\_type) where role\_id <= 7 {[}1.2M{]} \\
(cast\_info,role\_type) where role\_id <= 6 {[}1.1M{]} \\
(cast\_info,title) where kind\_id <= 4,production\_year <=\textbackslash \\ 2019,role\_id<= 6 {[}1.0M{]} \\
\end{tabular}                                                                   

& \begin{tabular}[c]{@{}l@{}}
(cast\_info,name) where role\_id \textless{}= 9 {[}2.1M{]}\\
(cast\_info,name) where role\_id\textless{}= 11 {[}2.1M{]}\\ (cast\_info,role\_type) where role\_id \textless{}= 11 {[}1.9M{]} \\
(cast\_info,role\_type) where role\_id \textless{}= 9 {[}1.8M{]} \\
(cast\_info,role\_type) where role\_id \textless{}= 7 {[}1.3M{]} \\
(cast\_info,name) where role\_id \textless{}= 8 {[}1.2M{]} \\
(cast\_info,role\_type) where role\_id \textless{}= 7 {[}1.1M{]} \\
(cast\_info,title) where kind\_id <= 4,production\_year <=\textbackslash \\ 2019,role\_id<= 6 {[}.9M{]} \\
\end{tabular}  \\ \hline
\multicolumn{1}{|c|}{DMV}   & \begin{tabular}[c]{@{}l@{}}

(accidents,time) where year\textless{}= 2005 and month \textless{}= 9 {[}11K{]}\\
(accidents,time) where month \textless{}= 9 and year \textless{}= 2005 {[}11K{]}\\
(accidents,time) where year \textless{}= 2003 and month \textless{}= 6 {[}10K{]} \\
(accidents,time) where year \textless{}= 2000 and month \textless{}= 6 {[}10K{]} \\
(accidents,location) where year \textless{}= 2001 {[}7K{]} \\
(accidents,location) where year \textless{}= 2000 {[}7K{]} \\
(car,accidents) where year \textless{}= 2005 {[}6K{]} \\
(car,accidents) where year \textless{}= 2004 {[}6K{]} 

\end{tabular}                               & \begin{tabular}[c]{@{}l@{}}

(accidents,time) where year \textless{}= 2005 and month \textless{}= 9 {[}33K{]}\\
(accidents,location) where year \textless{}= 2003 {[}18K{]}\\
(car,accidents) where year \textless{}= 2005 {[}18K{]} \\
(car,accidents) where year \textless{}= 2003 {[}18K{]} \\
(accidents,time) where year \textless{}= 2005 and month \textless{}= 9 {[}17K{]} \\
(accidents,time) where year \textless{}= 2003 and month \textless{}= 6 {[}10K{]} \\
(accidents,time) where year \textless{}= 2000 and month \textless{}= 6 {[}10K{]} \\
(accidents,location) where year \textless{}= 2002 {[}8K{]} \\
\end{tabular} \\ \hline

\multicolumn{1}{|c|}{TPC-H} & \begin{tabular}[c]{@{}l@{}}

(lineitem,orders) l\_linenumber \textless{}= 7 and l\_quantity \textless{}=16 {[}116K{]}\\
(lineitem,orders) l\_linenumber \textless{}= 6 and l\_quantity\textless{}= 16 {[}109K{]}\\
(lineitem,orders) l\_linenumber \textless{}= 5 and l\_quantity \textless{}= 35 {[}98K{]} \\
(lineitem,orders) l\_linenumber \textless{}= 7 and l\_quantity \textless{}= 34 {[}84K{]} \\
(lineitem,orders) l\_linenumber \textless{}= 7 and l\_quantity \textless{}= 27 {[}65K{]} \\
(lineitem,orders) l\_linenumber \textless{}= 2 and l\_quantity \textless{}= 17 {[}61K{]} \\
(lineitem,orders) l\_linenumber \textless{}= 6 and l\_quantity \textless{}= 27 {[}61K{]} \\
(lineitem,orders) l\_linenumber \textless{}= 7 and l\_quantity \textless{}= 23 {[}59K{]} 

\end{tabular} & 

\begin{tabular}[c]{@{}l@{}}

(lineitem,orders) where l\_linenumber \textless{}= 7 and l\_quantity \textless{}= 34 {[}89K{]}\\
(lineitem,orders) where l\_linenumber\textless{}= 6 and l\_quantity \textless{}= 38 {[}75K{]}\\
(lineitem,orders) where l\_linenumber \textless{}= 5 and l\_quantity \textless{}= 35 {[}72K{]} \\
(lineitem,orders) where l\_linenumber \textless{}= 6 and l\_quantity \textless{}= 28 {[}62K{]} \\
(lineitem,orders) where l\_linenumber \textless{}= 7 and l\_quantity \textless{}= 22 {[}55K{]} \\
(lineitem,orders) where l\_linenumber \textless{}= 5 and l\_quantity \textless{}= 28 {[}52K{]} \\
(lineitem,orders) where l\_linenumber \textless{}= 4 and l\_quantity \textless{}= 28 {[}51K{]} \\
(lineitem,orders) where l\_linenumber \textless{}= 7 and l\_quantity \textless{}= 23 {[}50K{]} 

\end{tabular} \\ \hline

\end{tabular}
}
\caption{The $2Join$ Queries with the Highest Errors from the NN and RNN Models: For each dataset, we show the top eight queries with the highest absolute errors.}
\label{tab:hard_2join}

\end{table*}

%% file: EvalFigure_Snippets/fig_column_distributions.tex
\begin{figure*}[t]
    \begin{subfigure}[b]{.18\linewidth}
        \includegraphics[width=\linewidth]{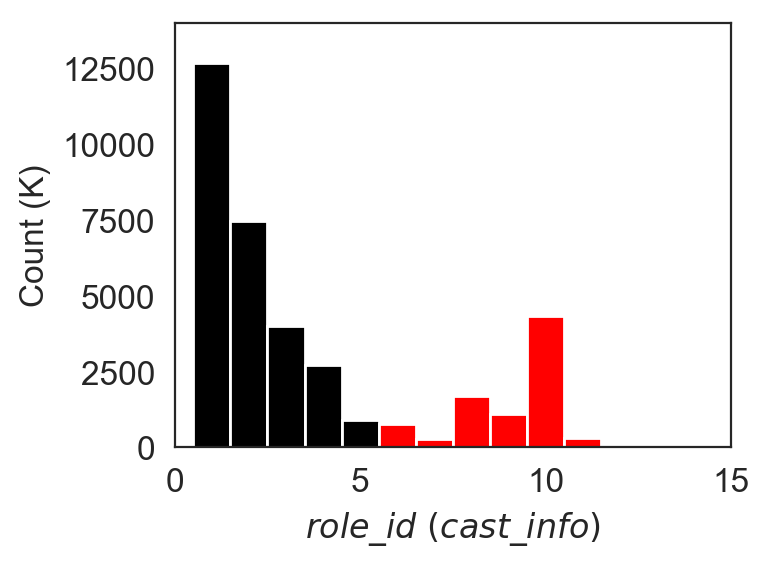}
        \caption{role\_id}
        \label{fig:roleid}
    \end{subfigure} %
    \begin{subfigure}[b]{.18\linewidth}
        \includegraphics[width=\linewidth]{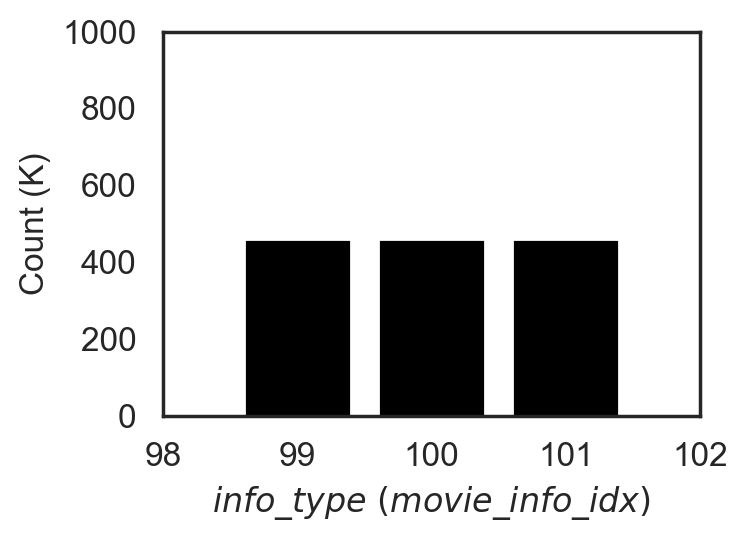}
        \caption{info\_type}
        \label{fig:infotype}
   \end{subfigure}
   \begin{subfigure}[b]{.18\linewidth}
        \includegraphics[width=\linewidth]{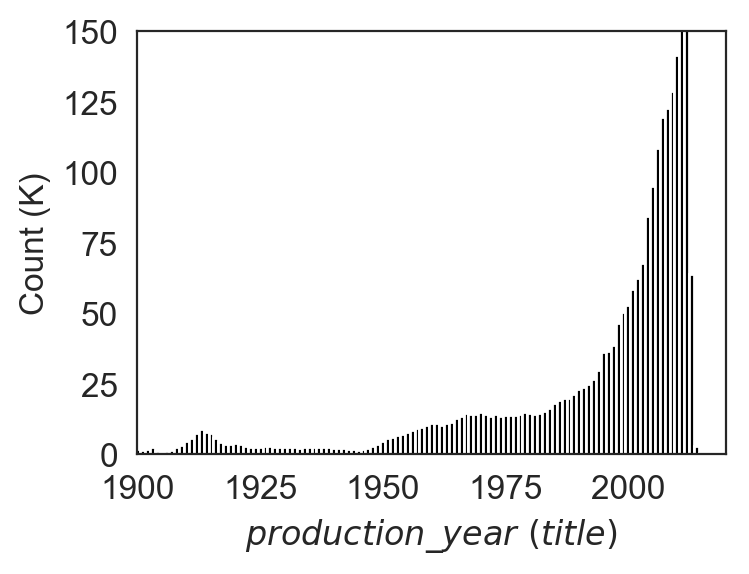}
        \caption{production\_year}
        \label{fig:productionyear}
   \end{subfigure}
   \begin{subfigure}[b]{.18\linewidth}
        \includegraphics[width=\linewidth]{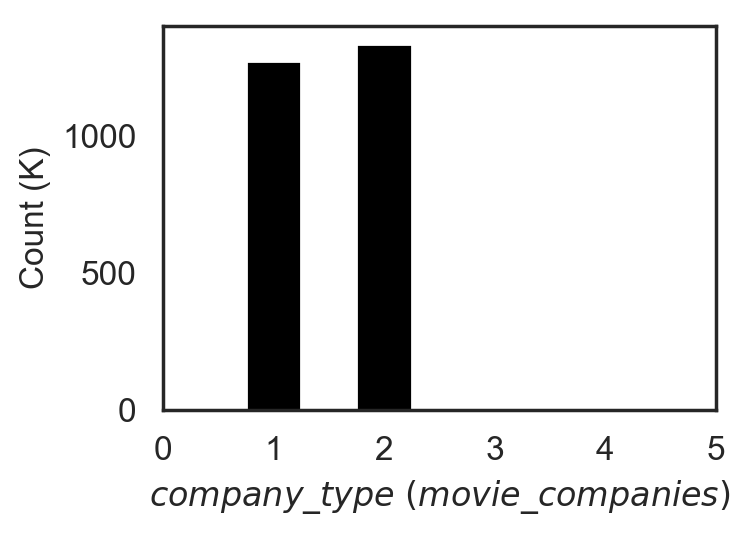}
        \caption{company\_type}
        \label{fig:companytype}
   \end{subfigure}
   \begin{subfigure}[b]{.18\linewidth}
        \includegraphics[width=\linewidth]{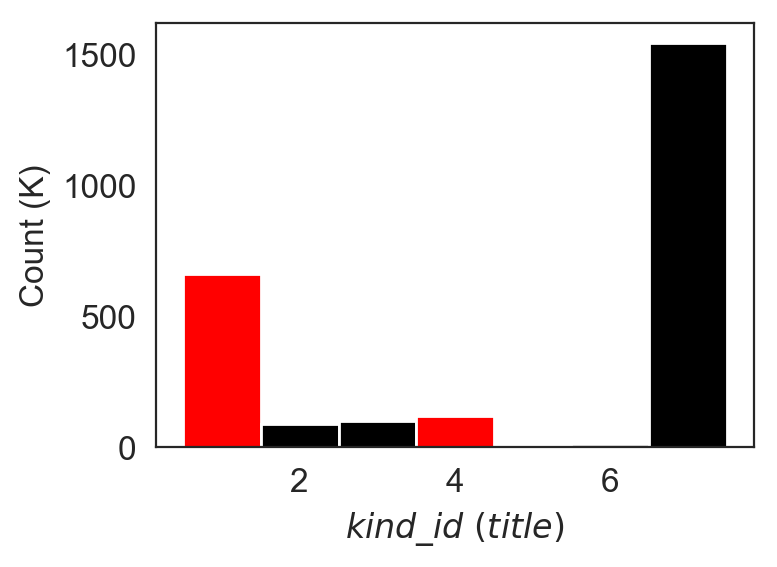}
        \caption{kind\_id}
        \label{fig:kindid}
   \end{subfigure}
   
   \bigskip

   \begin{subfigure}[b]{.18\linewidth}
        \includegraphics[width=\linewidth]{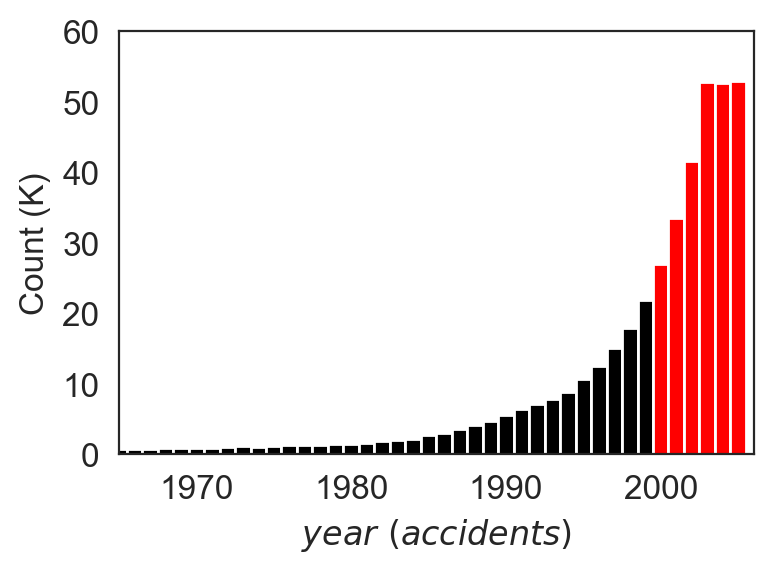}
        \caption{year (accidents)}
        \label{fig:year}
    \end{subfigure} %
    \begin{subfigure}[b]{.18\linewidth}
        \includegraphics[width=\linewidth]{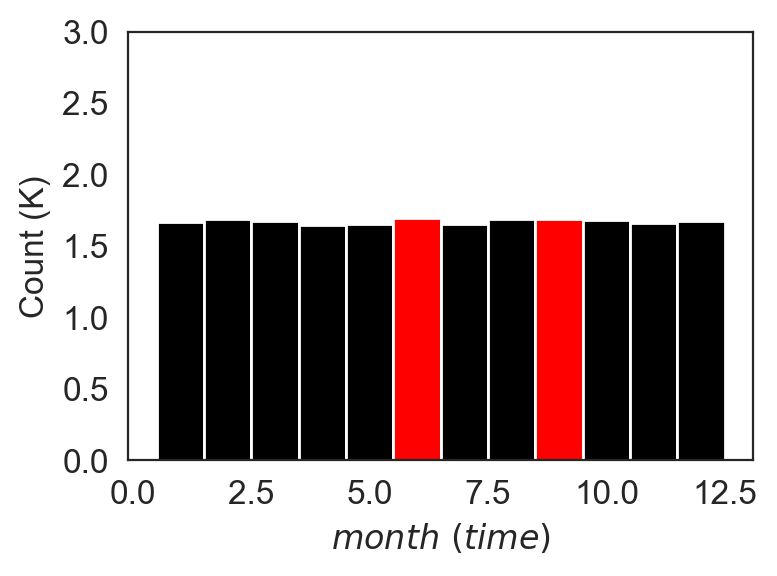}
        \caption{month}
        \label{fig:month}
   \end{subfigure}
   \begin{subfigure}[b]{.18\linewidth}
        \includegraphics[width=\linewidth]{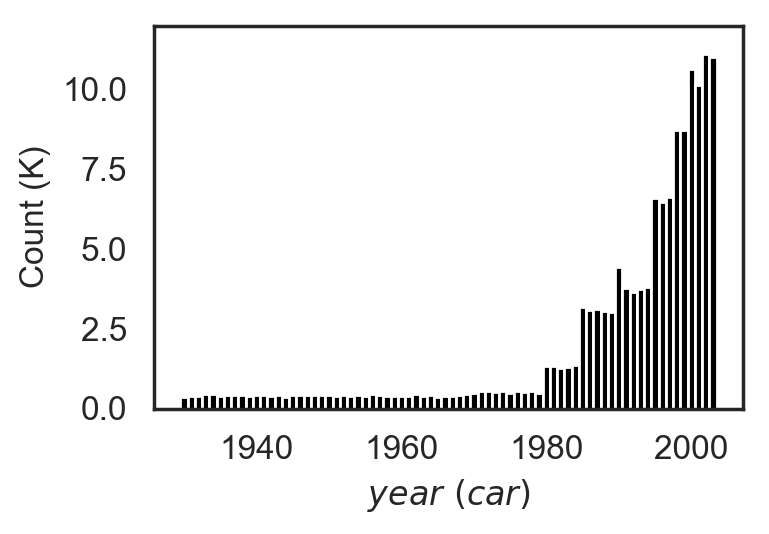}
        \caption{year (car)}
        \label{fig:yearcar}
   \end{subfigure}
   \begin{subfigure}[b]{.18\linewidth}
        \includegraphics[width=\linewidth]{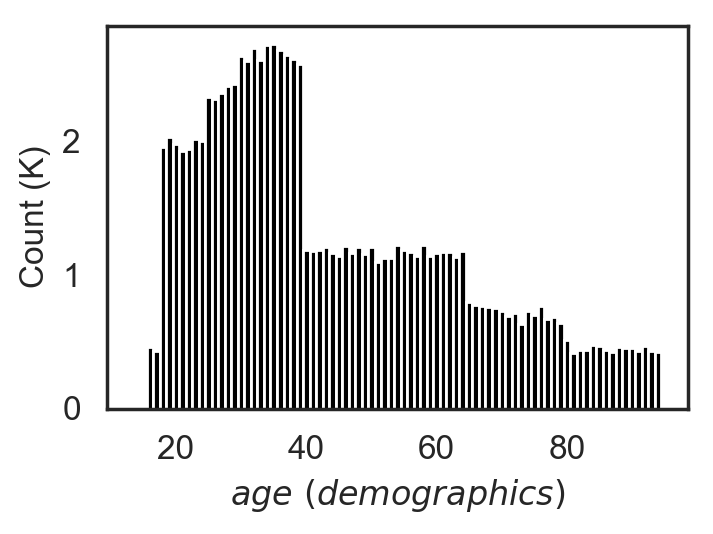}
        \caption{age}
        \label{fig:age}
   \end{subfigure}

   \bigskip
   
   \begin{subfigure}[b]{.18\linewidth}
        \includegraphics[width=\linewidth]{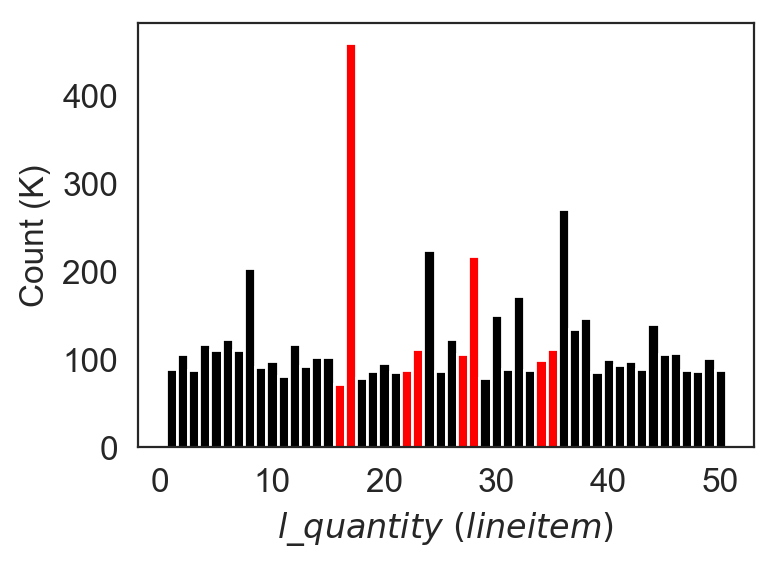}
        \caption{l\_quantity}
        \label{fig:lquantity}
    \end{subfigure} %
    \begin{subfigure}[b]{.18\linewidth}
        \includegraphics[width=\linewidth]{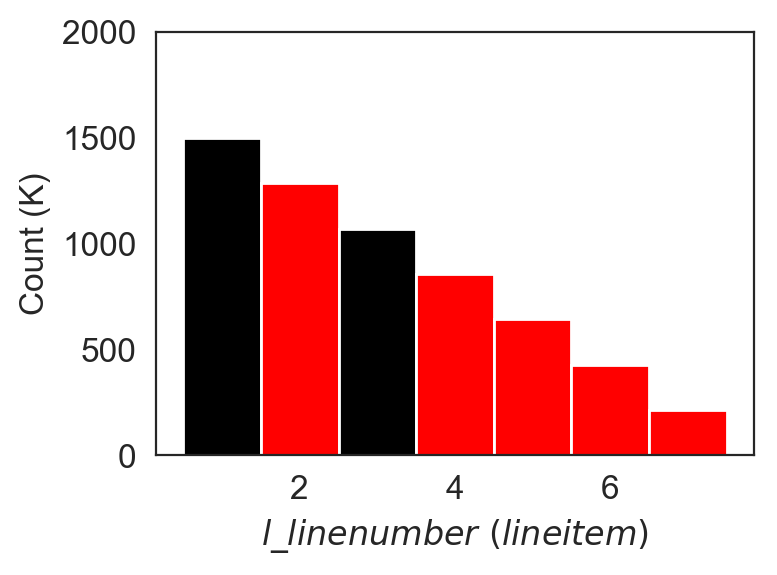}
        \caption{l\_linenumber}
        \label{fig:linenumber}
   \end{subfigure}
   \begin{subfigure}[b]{.18\linewidth}
        \includegraphics[width=\linewidth]{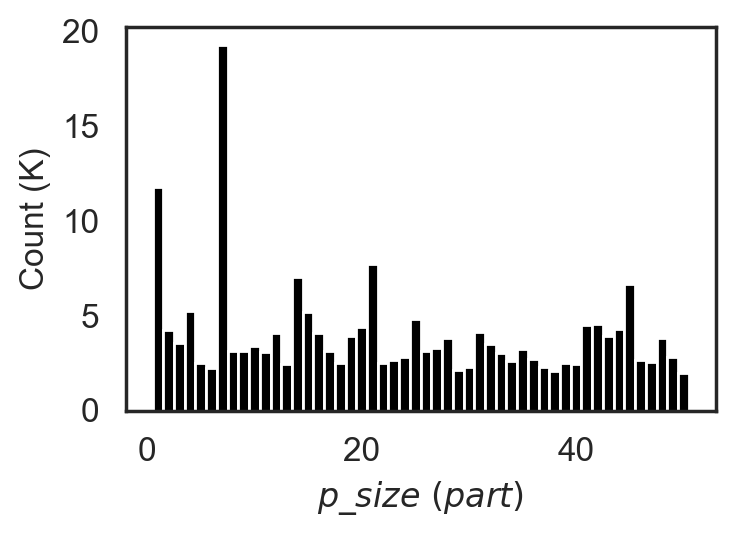}
        \caption{p\_size}
        \label{fig:psize}
   \end{subfigure}
   \begin{subfigure}[b]{.18\linewidth}
        \includegraphics[width=\linewidth]{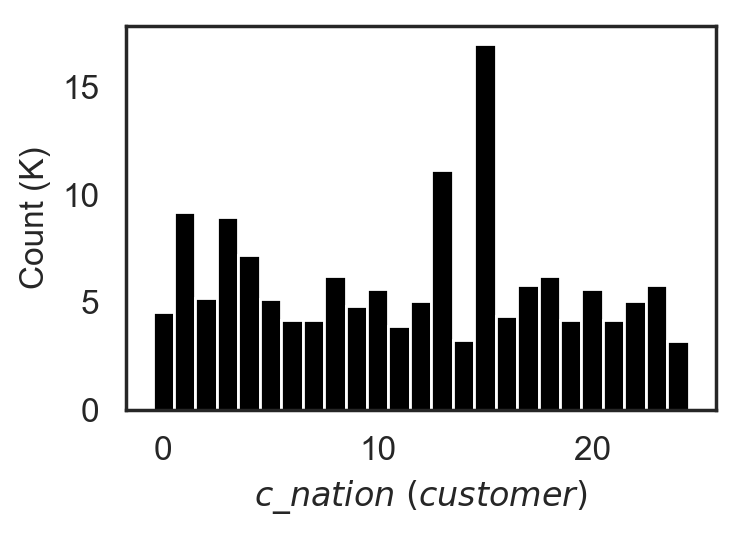}
        \caption{c\_nation}
        \label{fig:nation}
   \end{subfigure}

   \caption{Distributions for all Selection Columns: First row shows all distributions from the IMDB relation. Second shows distributions from DMV, and the third shows TPC-H.}
    \label{fig:column_distributions}
\end{figure*}

%% file: EvalFigure_Snippets/fig_cdf_unlimited_storage.tex
\begin{figure*}[t]

\begin{subfigure}{.28\linewidth}
\includegraphics[width=\linewidth]{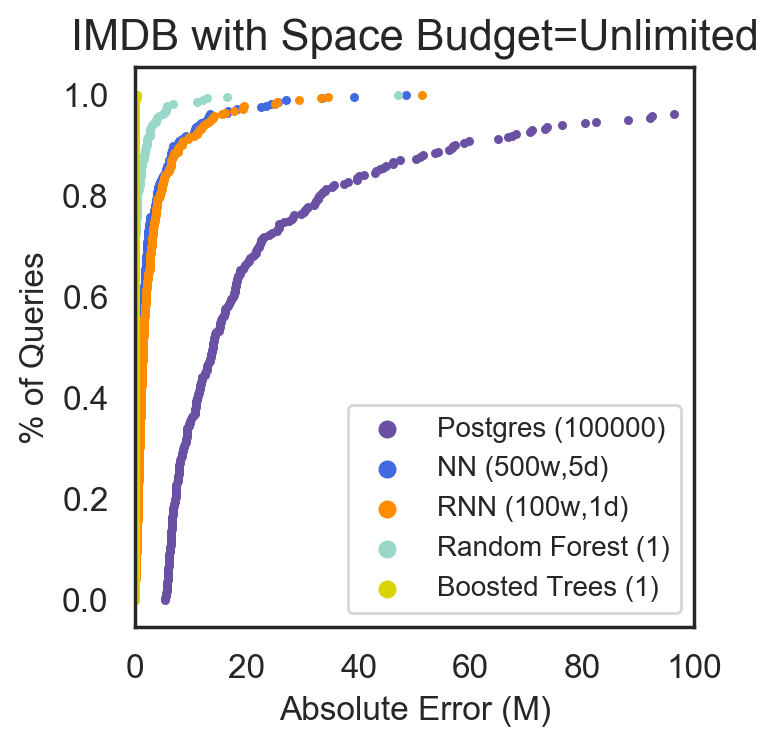}%
\caption{IMDB Unlimited}%
\label{fig:unlimited_imdb}%
\centering
\scalebox{0.7}{
\begin{tabular}{|l|lll|}
\hline
\multicolumn{4}{|c|}{CDF Percentiles}       \\ \hline
           & $25\%$ & $50\%$ & $75\%$ \\ \hline
PostgreSQL & 7.8M & 13.9M & 27.5M \\ 
NN (500w,5d)  & .30M  & .97M & 2.6M    \\
RNN (100w,1d) & .6M &  1.4M & 3.6M   \\ 
Random Forest (1) & 1e-6M & 6e-6M & .2M \\ 
Boosted Trees (1) & 8e-7M & 2e-6M & 8e-6M \\ \hline
\end{tabular}
}

\bigskip

\scalebox{0.7}{
\begin{tabular}{|l|lll|}
\hline
\multicolumn{4}{|c|}{Average Absolute Errors}       \\ \hline
         & $2Join$ & $4Join$ & $6Join$ \\ \hline
PostgreSQL & 6.8M & 12.8M & 31.3M \\ 
NN (500w,5d)  & .28M  & .45M & 4.1M    \\
RNN (100w,1d) & .58M &  2.0M & 4.1M  \\ 
Random Forest (1) & 1e-6 & .21M & 1.0M \\ 
Boosted Trees (1) & 1e-6 & 1e-4 & 7e-4 \\ \hline
\end{tabular}
}

\bigskip

\scalebox{0.7}{
\begin{tabular}{|l|lll|}
\hline
\multicolumn{4}{|c|}{Average Relative Errors}       \\ \hline
           & $2Join$ & $4Join$ & $6Join$ \\ \hline
PostgreSQL & .38 & .80 & .95 \\ 
NN (500w,5d)  & .01  & .03 & .11    \\
RNN (100w,1d) & .03 &  .11 & .13   \\ 
Random Forest (1) & 3e-8 & .01 &  .03 \\ 
Boosted Trees (1) & 3e-8 & 9e-5 & 1e-4 \\ \hline
\end{tabular}
}

\end{subfigure}
\begin{subfigure}{.28\linewidth}
\includegraphics[width=\linewidth]{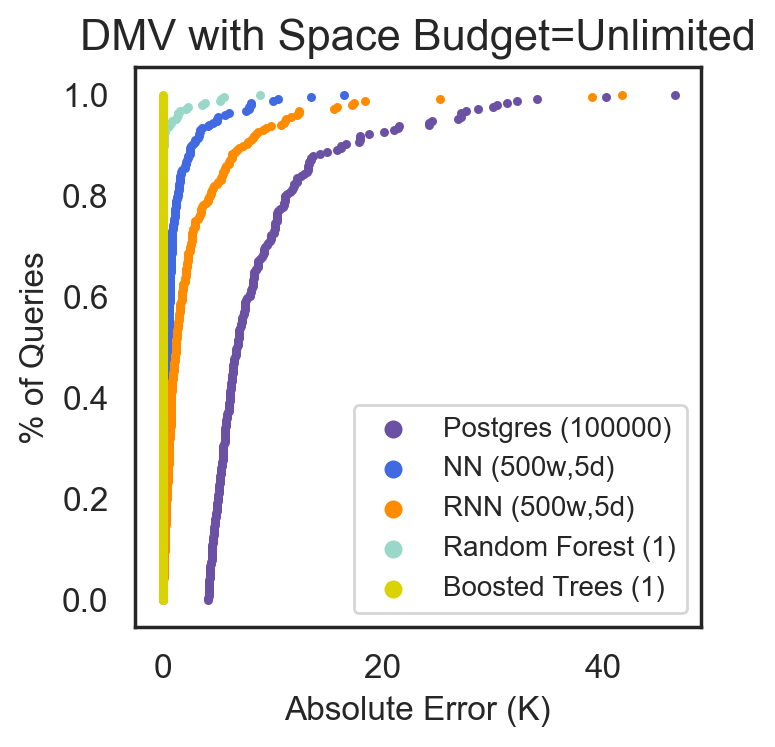}%
\caption{DMV Unlimited}%
\label{fig:unlimited_dmv}%
\centering
\scalebox{0.7}{
\begin{tabular}{|l|lll|}
\hline
\multicolumn{4}{|c|}{CDF Percentiles}       \\ \hline
           & $25\%$ & $50\%$ & $75\%$ \\ \hline
PostgreSQL &  5.28K & 6.82K  & 10.3K \\ 
NN (500w,5d)  & .16K & .46K & 1.0K   \\
RNN (500w,5d) & .02K & 1.1K & 3.2K    \\ 
Random Forest (1) & 3e-6K & 6e-6K & 1e-5K   \\ 
Boosted Trees (1) & 3e-6K & 5e-6K & 1e-5K \\ \hline
\end{tabular}
}

\bigskip

\scalebox{0.7}{
\begin{tabular}{|l|lll|}
\hline
\multicolumn{4}{|c|}{ Average Absolute Errors}       \\ \hline
         & $2Join$ & $4Join$ & $6Join$ \\ \hline
PostgreSQL & 9.5K & 9.3K & 8.9K \\ 
NN (500w,5d)  &  .4K  & .9K & 1.3K   \\
RNN (500w,5d) &  1.1K & 3.3K  & 2.8K   \\ 
Random Forest (1) & 8e-5 & .06K & .29K  \\ 
Boosted Trees (1) & 8e-5 & 2e-3 & 9e-3 \\ \hline
\end{tabular}
}

\bigskip

\scalebox{0.7}{
\begin{tabular}{|l|lll|}
\hline
\multicolumn{4}{|c|}{ Average Relative Errors}       \\ \hline
           & $2Join$ & $4Join$ & $6Join$ \\ \hline
PostgreSQL & .10 & .20 &  .23 \\ 
NN (500w,5d)  & .002 & .01 & .01     \\
RNN (500w,5d) & .006 & .03 & .04  \\ 
Random Forest (1) & 7e-8 & .0008 & .007  \\ 
Boosted Trees (1) & 7e-8 & 6e-6 & 4e-5 \\ \hline
\end{tabular}
}
\end{subfigure}
\begin{subfigure}{.28\linewidth}
\includegraphics[width=\linewidth]{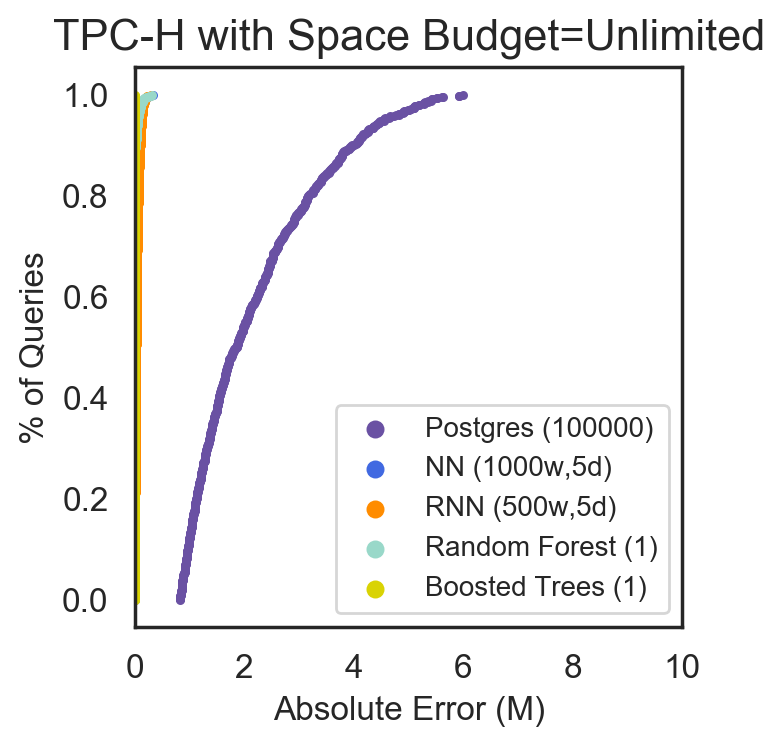}%
\caption{TPCH Unlimited}%
\label{fig:unlimited_tpch}

\centering
\scalebox{0.7}{
\begin{tabular}{|l|lll|}
\hline
\multicolumn{4}{|c|}{CDF Percentiles}       \\ \hline
           & $25\%$ & $50\%$ & $75\%$ \\ \hline
PostgreSQL & 1.2M & 1.8M & 2.9M \\ 
NN (1000w,5d)  & .01M & .02M & .04M    \\
RNN (500w,5d) & .01M & .02M & .05M   \\ 
Random Forest (1) & 4e-8M & 8e-8M & 1e-7M  \\
Boosted Trees (1) & 3e-8M & 7e-8M & 1e-7M \\ \hline
\end{tabular}
}

\bigskip

\scalebox{0.7}{
\begin{tabular}{|l|lll|}
\hline
\multicolumn{4}{|c|}{Average Absolute Errors}       \\ \hline
         & $2Join$ & $4Join$ & $6Join$ \\ \hline
PostgreSQL & 2.9M & 2.2M & 1.8M \\ 
NN (1000w,5d)  & 16K   & 41K & 28K  \\
RNN (500w,5d) & 29K & 59K & 23K   \\ 
Random Forest (1) & 9e-4 & 3K & 15K   \\ 
Boosted Trees (1) & 9e-4 & 1e-3 & 4e-2 \\ \hline
\end{tabular}
}

\bigskip

\scalebox{0.7}{
\begin{tabular}{|l|lll|}
\hline
\multicolumn{4}{|c|}{Average Relative Errors}       \\ \hline
           & $2Join$ & $4Join$ & $6Join$ \\ \hline
PostgreSQL & .99 & .99 & .99 \\ 
NN (1000w,5d)  & .007 & .02 & .01    \\
RNN (500w,5d) & .01 & .02 & .01  \\ 
Random Forest (1) & 4e-8 & .001 & .009  \\
Boosted Trees (1) & 4e-8 & 9e-8 & 4e-6 \\ \hline
\end{tabular}
}
\end{subfigure}

\caption{Error Analysis for all Models : We show the curve for Hard(PostgreSQL) and show the corresponding errors from the best models with an unlimited storage budget. Below each graph, we show tables detailing the percentiles, the average absolute error and average relative error.}
\label{fig:cdf_unlimited_storage}%

\end{figure*}

%% file: EvalFigure_Snippets/fig_trade-offs.tex
\begin{figure*}[t]
    \centering
    \begin{subfigure}[b]{.9\linewidth}
        \centering
        \includegraphics[width=\linewidth]{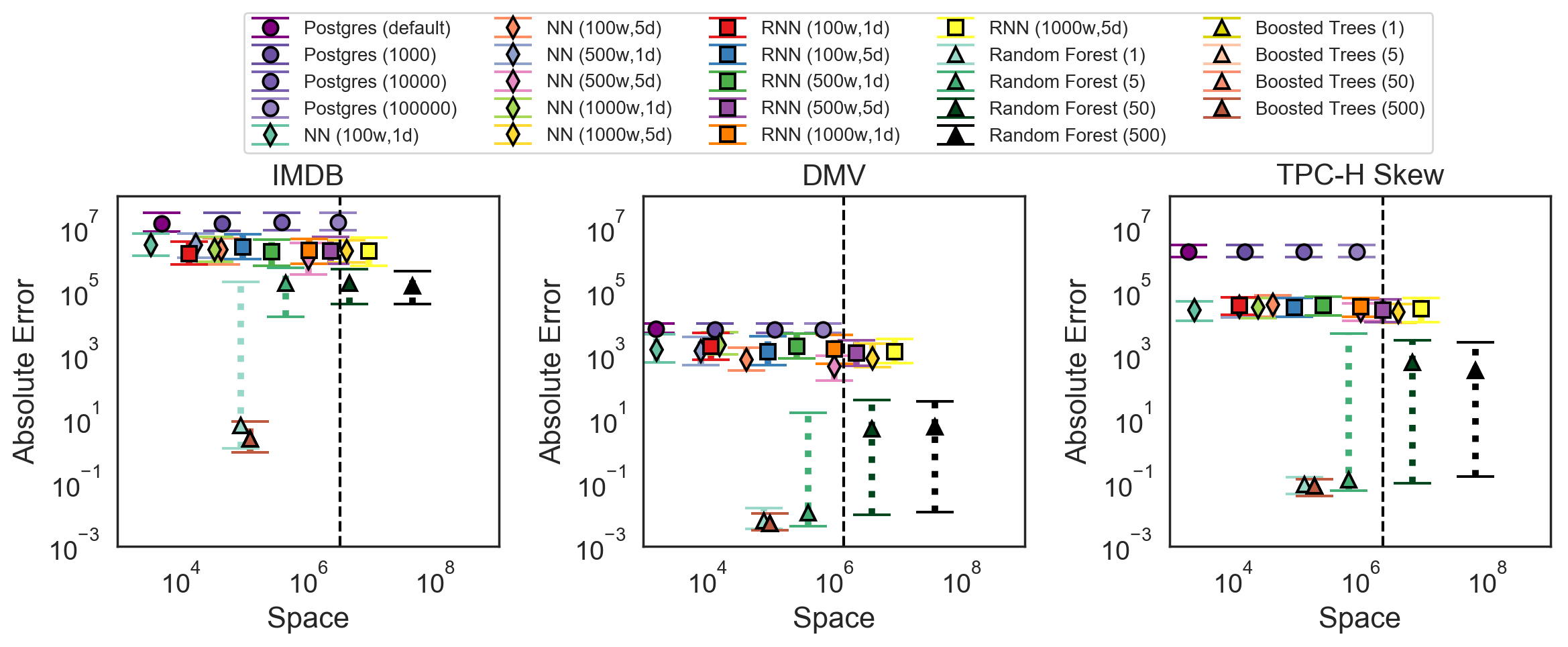}
        \caption{ Space vs. Error for all models }
        \label{fig:space}
    \end{subfigure} %
    \centering
    \begin{subfigure}[b]{.9\linewidth}
        \centering
        \includegraphics[width=\linewidth]{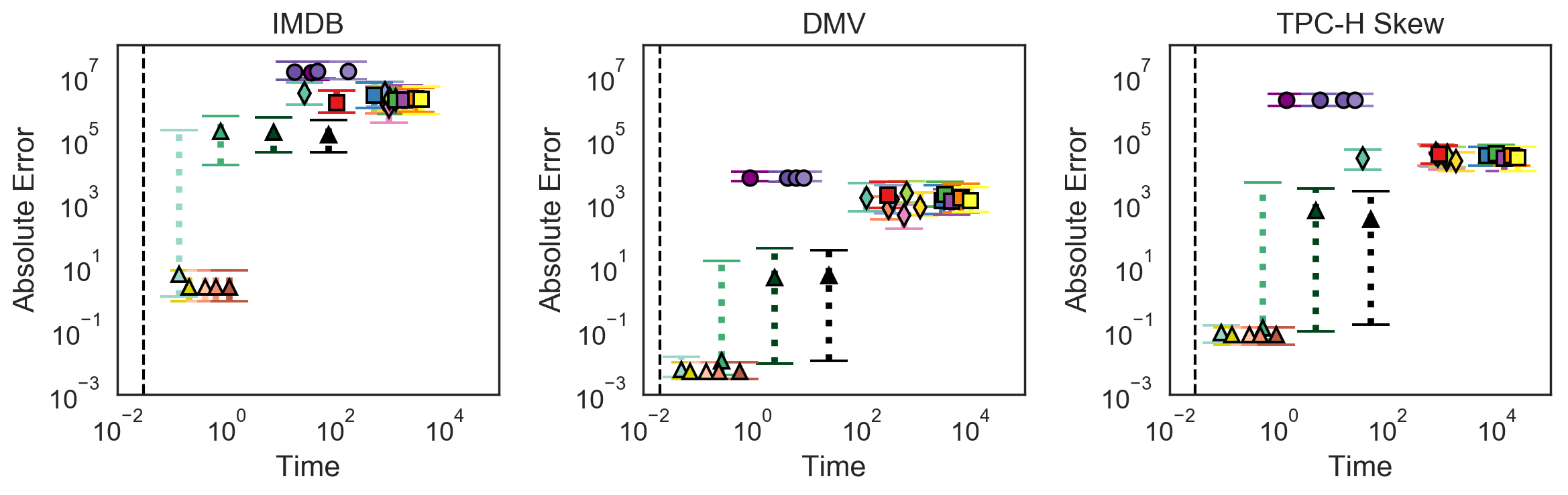}
        \caption{ Time vs Error for all models}
        \label{fig:time}
    \end{subfigure}
    \caption{Trade-offs between Error, Space and Time: We show the absolute error, space and time for each model and for PostreSQL for different number of bins. The horizontal line represents the space and time for the hash table model.}
    \vspace{.5cm}
    \label{fig:trade-offs}
\end{figure*}

%% file: EvalFigure_Snippets/fig_robust_selections.tex
\begin{figure*}[!htbp]
\centering

\begin{subfigure}{.33\linewidth}
\includegraphics[width=.9\linewidth]{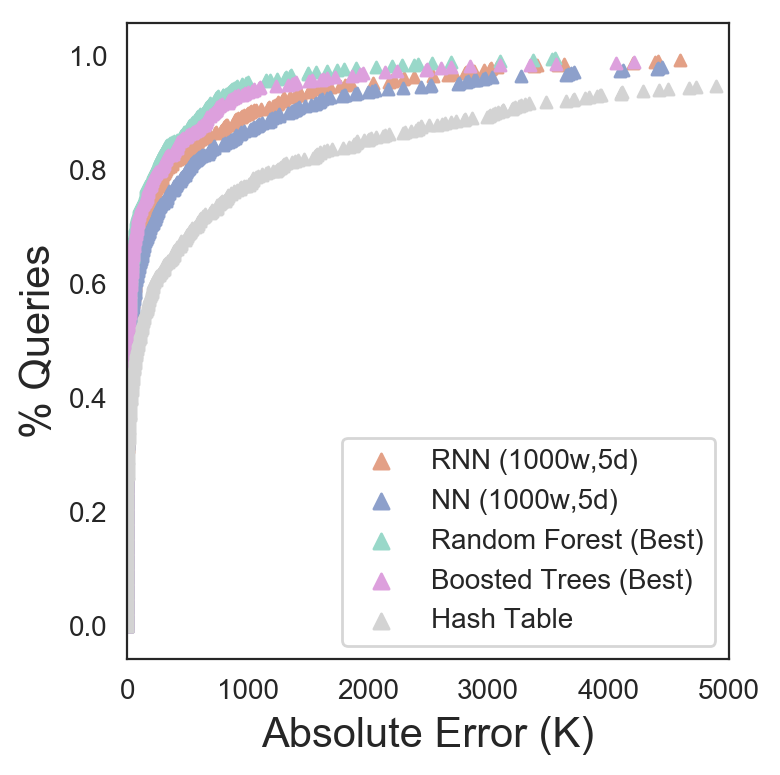}%
\caption{IMDB Remove Values \newline $production\_{year}$}%
\label{fig:im_remove_selection_1}
\end{subfigure}
\begin{subfigure}{.33\linewidth}
\includegraphics[width=.9\linewidth]{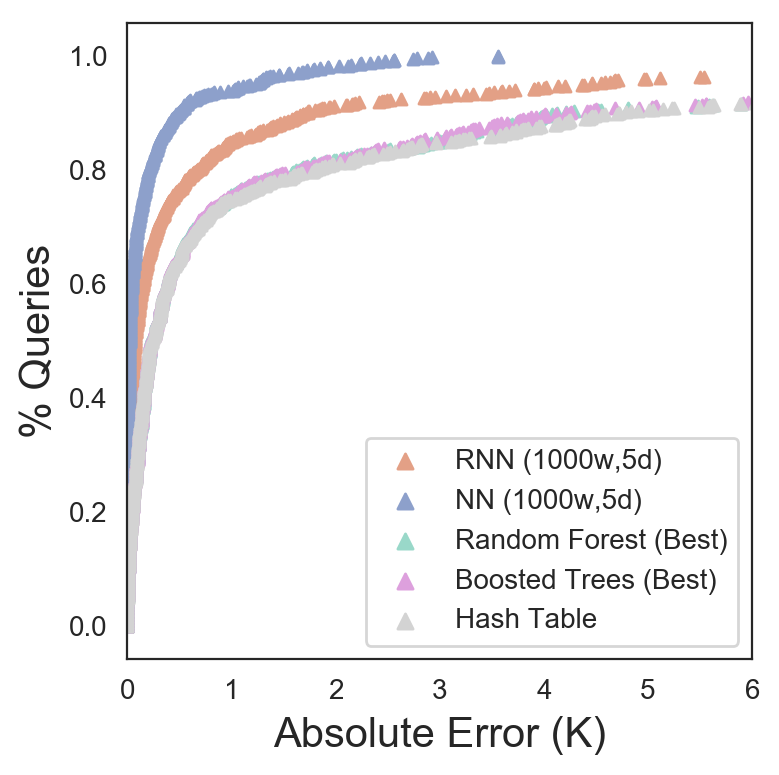}%
\caption{DMV Remove Selection Values \newline $year$}%
\label{fig:d_remove_selection_1}
\end{subfigure}
\begin{subfigure}{.33\linewidth}
\includegraphics[width=.9\linewidth]{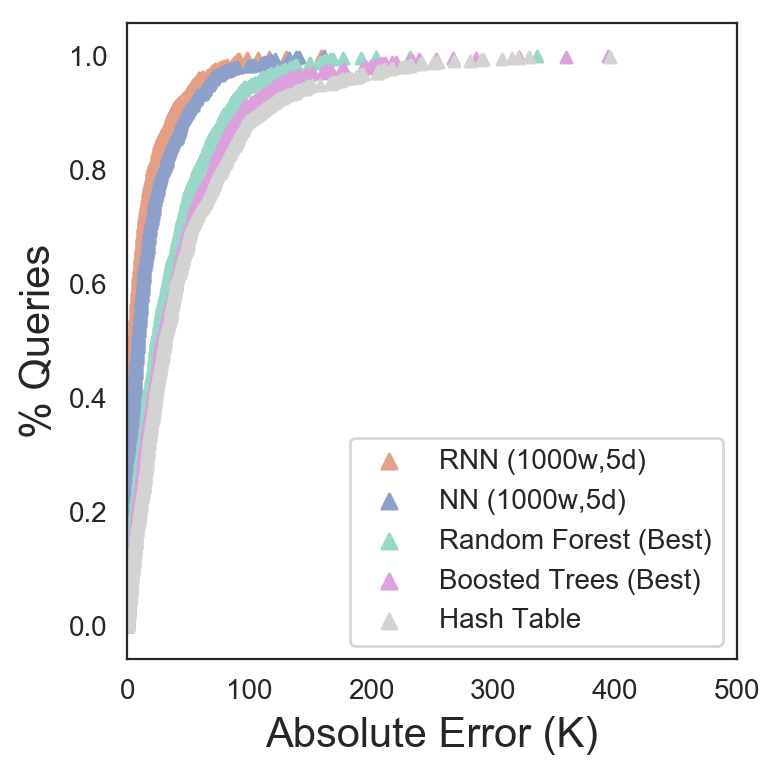}%
\caption{TPC-H Remove Selection \newline $l\_{quantity}$}%
\label{fig:t_remove_selection_1}
\end{subfigure}

\caption{Removing 10\% selection predicate values across all datasets}
\label{fig:robust_selections}

\end{figure*}

%% file: EvalFigure_Snippets/fig_robust_joins.tex
\begin{figure}[t]
\centering
\begin{subfigure}{.7\linewidth}
\includegraphics[width=.9\linewidth]{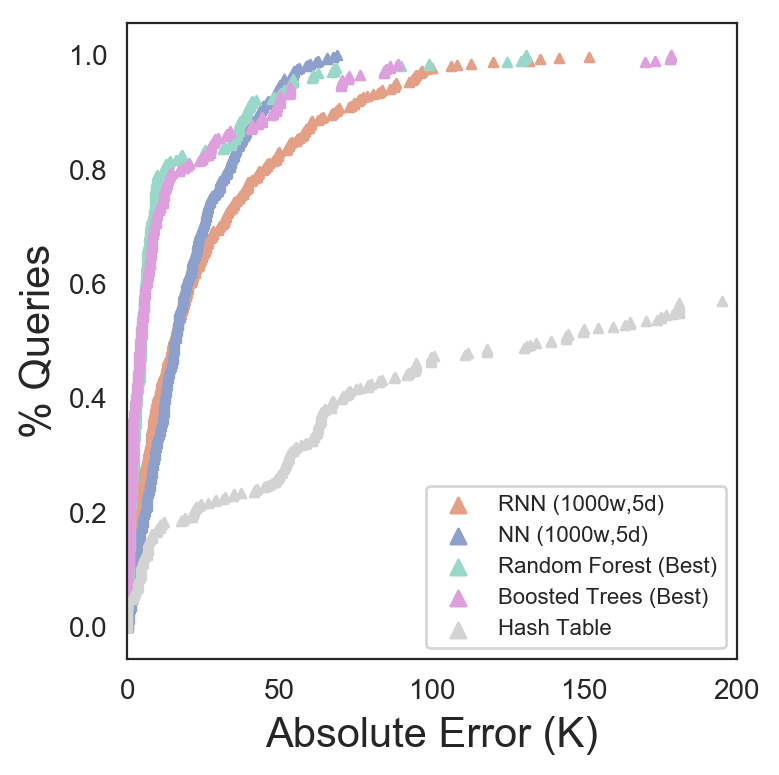}
\caption{IMDB Remove Join}
\label{fig:im_remove_join}
\end{subfigure}
\begin{subfigure}{.7\linewidth}
\includegraphics[width=.9\linewidth]{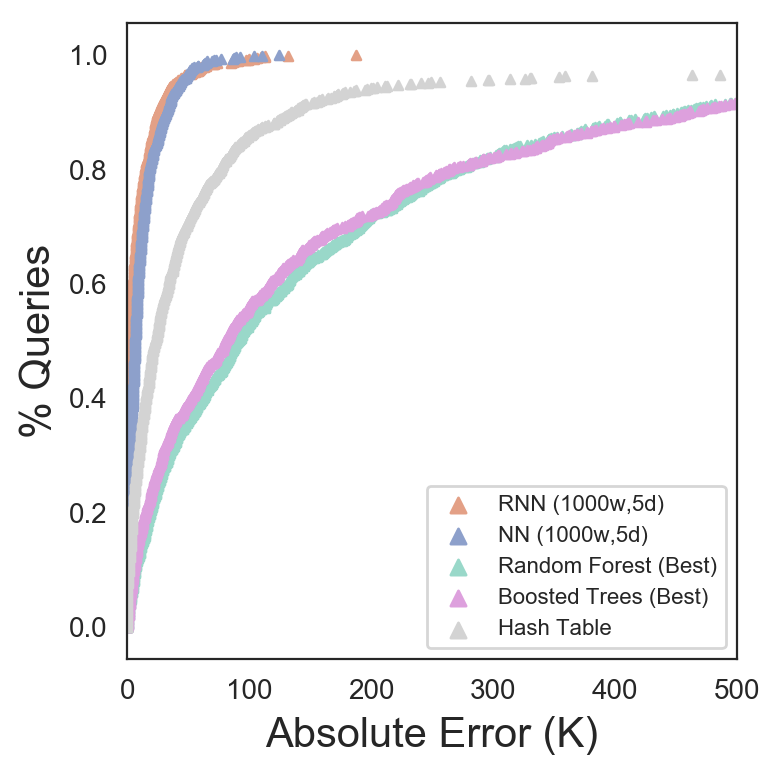}
\caption{TPC-H Remove Join}
\label{fig:t_remove_join}
\end{subfigure}
\caption{Removing joins from the training workload}
\label{fig:robust_joins}
\end{figure}

%% file: EvalFigure_Snippets/fig_latent_representations.tex
\begin{figure}[t]
\includegraphics[width=\linewidth]{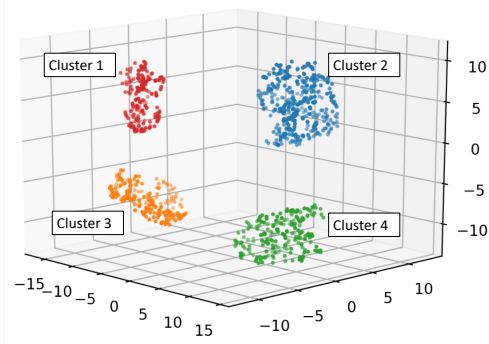}
\vspace{-1.5cm}
\caption{Clustering dimensionality-reduced latents for the NN (100w, 1d) model on the TPC-H dataset.}
\label{fig:latent_representations}
\end{figure}

%% file: EvalFigure_Snippets/fig_speed_up.tex
\begin{figure*}%
\centering
\begin{subfigure}{.30\textwidth}
\includegraphics[width=\linewidth]{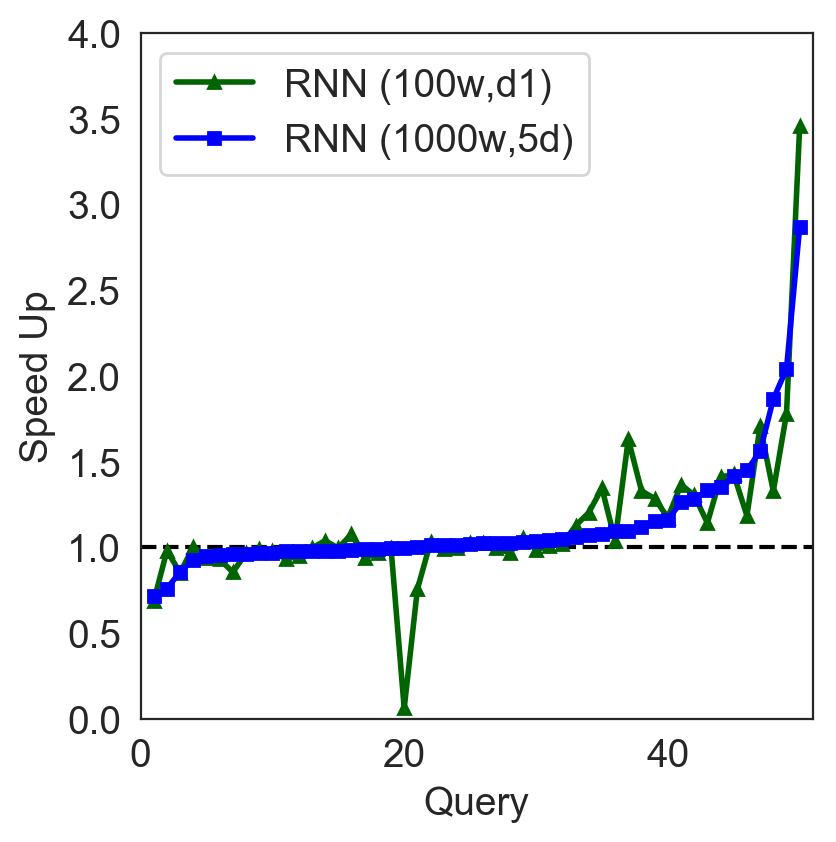}
\caption{IMDB Impact}
\label{fig:imdb_impact}
\end{subfigure}%
\begin{subfigure}{.30\textwidth}
\includegraphics[width=\linewidth]{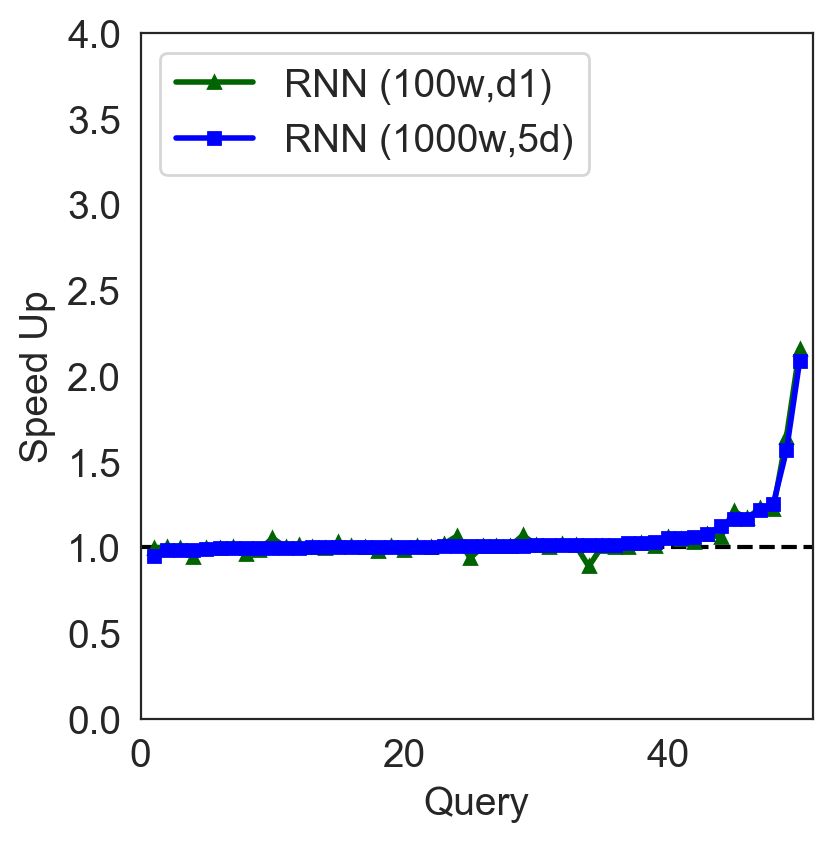}
\caption{DMV Impact}
\label{fig:dmv_impact}
\end{subfigure}%
\begin{subfigure}{.30\textwidth}
\includegraphics[width=\linewidth]{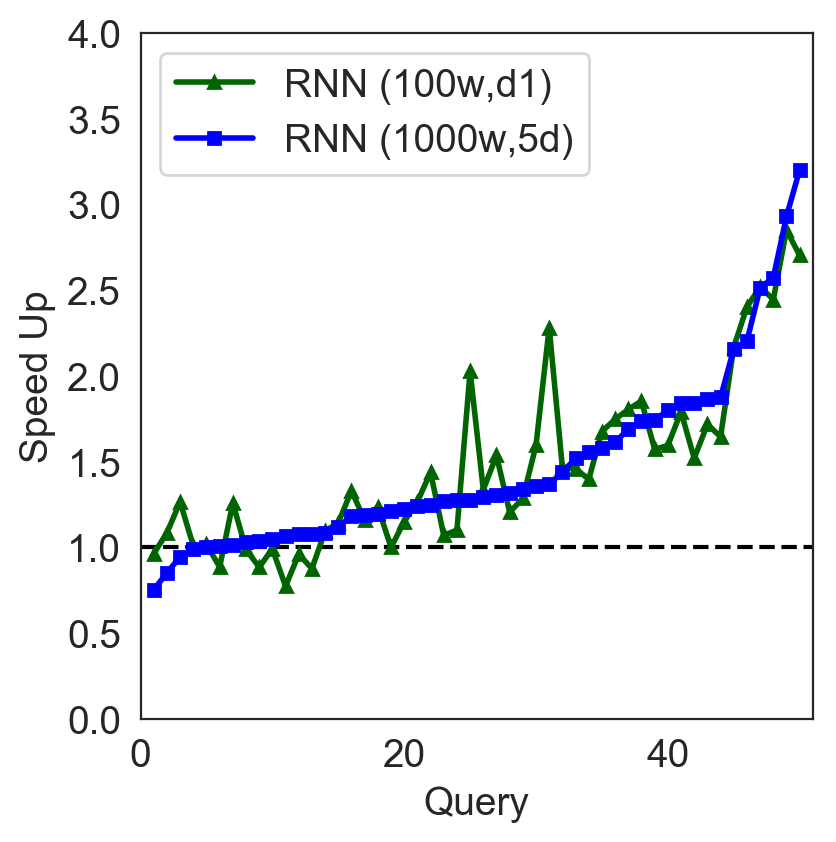}
\caption{TPC-H Impact}
\label{fig:tpch_impact}
\end{subfigure}%

\caption{Query execution time speed-ups thanks to cardinality estimates from simple or complex RNN}

\label{fig:speed_up}
\end{figure*}

%% file: practical_considerations.tex
\vspace{1cm}
\section{Practical Considerations}~\label{sec:practical}

In this section, we study two additional practical considerations. In \autoref{sec:measurement_analysis}, we evaluated the accuracy of cardinality estimates produced by the different models. In this section, we evaluate whether the cardinality estimate improvements lead to faster query execution plans. Additionally, in earlier sections, we showed the trade-offs between prediction error and time overhead due to model training. We did not consider the time that it takes to execute the training queries. To minimize this overhead, we discuss and consider using active learning as a way to reduce the time spent generating training sets.

\subsection{Impact on Query Plans}

We evaluate the impact of these models starting with a simple RNN model (100w, 1d) and going to a more complex one (1000w, 5d). We use the RNN, as query optimization requires evaluating cardinalities for several possible subqueries that could exist in the final plan. We evaluate the performance benefit for queries with 6 relations for each of the three datasets. As we collect the subquery cardinalities from the RNN, these estimates are then fed into a version of PostgreSQL modified to accept external cardinality estimates~\cite{Cai:19}. 

In \autoref{fig:speed_up}, we show the performance impact of these improved cardinalities compared to the default cardinality estimates from PostgreSQL. First, for the IMDB dataset, we show the performance improvement for 50 queries in \autoref{fig:imdb_impact}. The runtimes for these queries range from <1sec up to 200sec. The simple RNN model improves the performance of 54\% of the queries, while the complex model improves 60\% of the queries. For the simpler model, query 22 is an outlier where the model's estimates actually slows down the query considerably (from 2 seconds up to 39 seconds).  In contrast, there is no significant slow down on any query for the complex model. 

For the DMV dataset, both the simple model and complex model improve the performance for 76\% of the queries and there is no significant slow down for any query. We should note, however, that a majority of the query runtimes in this dataset range from 1 to 3 seconds. Finally, for the TPC-H dataset, the complex model improves 90\% of the queries. The simpler model also makes a significant improvement, speeding up 84\% of the queries. The query execution times for this dataset range from 20 to 120 seconds.

\subsection{Reducing the Training Time}

Building a model can be time consuming depending on the size of the model, the training time, and the amount of time that it takes to collect the training samples. To train the models shown in \autoref{sec:measurement_analysis}, we needed to run a large set of random queries to collect their ground-truth cardinalities, the output $Y$, for the models. Depending on the complexity of these queries, running them and collecting these labels can be time consuming. This process can be parallelized, but it comes with a resource cost. 

Models can be trained in several ways. One approach to reducing the time to collect training samples, is to train the model in an online fashion. That is, as the user executes queries while using the system, the model can train on only those queries. The learning happens in an incremental fashion, and updates the model after observing a batch of samples. This approach can work well if the user executes similar queries. Online learning can also be fast and memory efficient, but the learning may experience a drift~\cite{Gama:14}, where the model's decision boundary changes largely depends on the latest samples it observes. 


Alternatively, instead of relying on a user to provide query samples, we can use a technique known as \textit{active learning}. Active learning selects the best sample of candidates to improve a model's objective and to train as effectively as possible~\cite{Konyushkova:17}. It is ideal in settings where labeled examples are expensive to obtain~\cite{Cai:13}. 

Active learning works through a series of iterations. In each iteration, it determines unlabeled points to add to the training sample to improve the model. Given a large pool of unlabeled samples, active learning will select the unlabeled sample that should be annotated to improve the model's predictions. In our context, given a large pool of unlabeled queries, active learning should help narrow down which queries to execute next. 

There are various existing active learning methods. Common techniques include using uncertainty sampling, query-by-committee (QBC), and expected model change~\cite{Konyushkova:17}. In this work, we focus on using QBC~\cite{Seung:92}. After each active learning iteration, QBC first builds a committee of learners from the existing training dataset via bootstrapping~\cite{Wu:18a}. Each learner in the committee makes a prediction for all the samples in the unlabeled pool. The sample with the highest disagreement is labeled and added to the training pool. For regression tasks, this disagreement can be measured by the variance in the predictions across the learners~\cite{Repicky:2017}. 

Traditionally, active learning only adds a single informative sample in each data sampling iteration~\cite{Cai:17}. More recently, \textit{batch-model} AL (BMAL), where multiple samples are labeled in each iteration has become more prevalent, as labeling in bulk or in parallel has been more accessible in recent years~\cite{Chakraborty:15}. As shown in work by Wu et. al.~\cite{Wu:18a} careful attention must be placed in picking out diverse points with BMAL, as models might disagree on a batch that contains very similar points, leading to suboptimal results. 

We use BMAL in the following experiment and run three different methods to help select the unlabeled points for each iteration: 

\begin{enumerate}
    \item \textbf{QBC}: after each iteration, we train an ensemble of models and select the top $K$ points with the highest disagreement
    \item \textbf{QBC+Clustering}: we train an ensemble of models, but pick out the top $K$ \textit{diverse} set of points through clustering, which is based on the technique from \cite{Wu:18a} for linear regression
    \item \textbf{Random}: we select a random sample of points from the unlabeled pool
\end{enumerate}

\input{EvalFigure_Snippets/fig_active_learning.tex}

For each dataset, we use training samples from the $2Join$, $4Join$, and $6Join$ set along with all their subqueries, for a total of 600K samples for the model. We run two experiments. In the first experiment, we start with a small number of training samples (100) and set $K$=100. For the second experiment, we start with a larger sample (1000) and set $K$=1000. As the number of training samples is small, we include regularization to prevent overfitting.

In \autoref{fig:active_learning}, we show the loss of each technique for three active learning iterations on each dataset. We show the results for both experiments ($K$=100 and $K$=1000). Each point represents the average loss for three separate runs. For each graph, we also include the loss for the case where all samples are labeled (labeled as ``all training'').

In general, we find that with small training sets, QBC and QBC+Clustering result in a lower loss, particularly at the end of the first iteration. For subsequent iterations, the random technique performs just as well and in some cases even better, as in the TPC-H dataset for example. QBC is competitive, but it often overfits as shown by the cases where the loss increases (IMDB and TPC-H). This is expected, as BMAL techniques are known to select a distinct set of points to improve the loss more effectively. 

When the training set is larger ($K$=1000), all techniques perform similarly, negating the immediate benefit of active learning. Nevertheless, adding fewer points rather than the entire training set can still reach a loss that is approximately an order of magnitude away from the loss that includes all the training data. 

%% file: EvalFigure_Snippets/fig_active_learning.tex
\begin{figure*}[t]
\includegraphics[width=.9\linewidth]{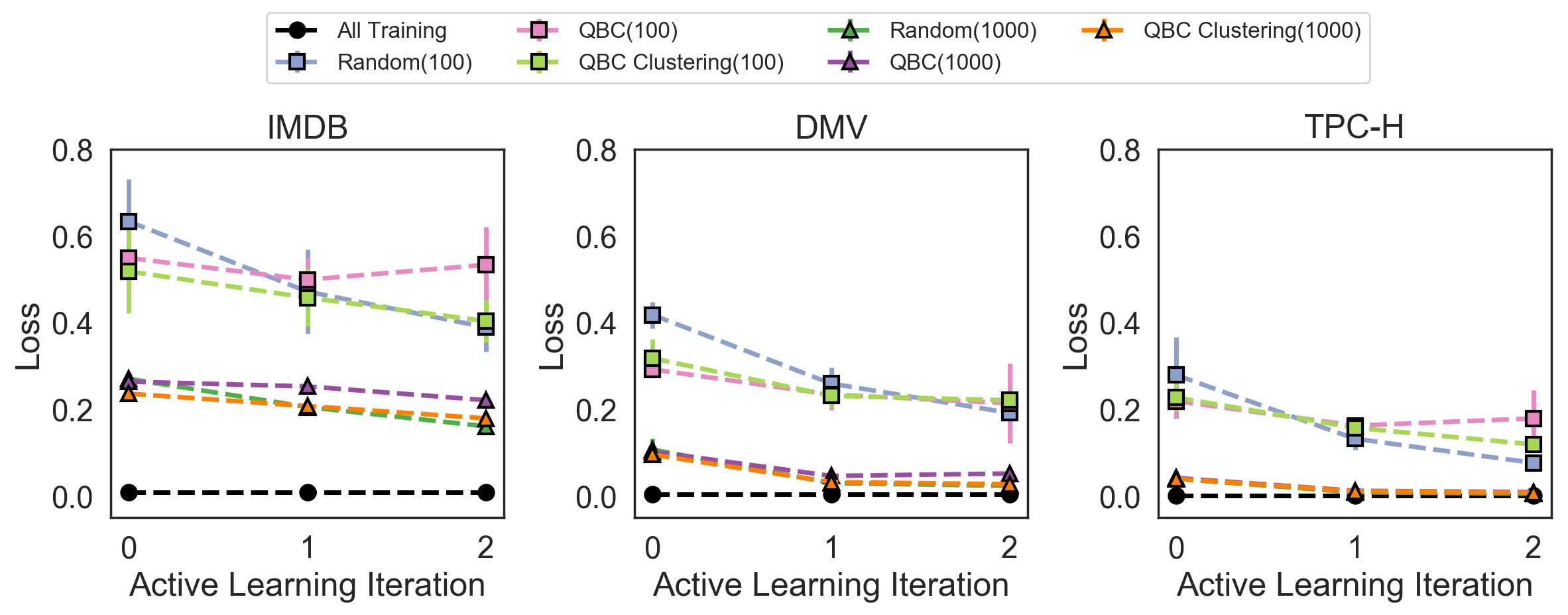}
\caption{Active Learning}
\label{fig:active_learning}
\end{figure*}

%% file: related_work.tex
\section{Related Work}\label{sec:relatedwork}

\textbf{Learning Optimizers}
Leo~\cite{Stillger:01}, was one of the first approaches to automatically adjust an optimizer's estimates based on past mistakes. This requires successive runs of similar queries to make adjustments. 

Similarly, in the effort of using a self-correcting loop, others have proposed a ``black-box'' approach to cardinality estimation by grouping queries into syntactic families~\cite{malik:07}. Machine learning techniques are then used to learn the cardinality distributions of these queries based on features describing the query attributes, constants, operators and aggregates. They specifically focus on applications that have fixed workloads do not require fine-grained, sub-plan estimates.

Work by Marcus \ea~\cite{Marcus:19} uses a deep reinforcement learning technique to find optimal join orders to improve query latency on a fixed database. They use cost estimates from PostgreSQL to bootstrap the learning and continuously improve the accuracy of the model's rewards during training. Related work by Sanjay \ea~\cite{Krishnan:19}, also uses deep reinforcement learning to improve query plans, but they assume perfect cardinality predictions for base relations. 

\textbf{Neural Networks and Cardinality Estimation}
Liu \ea~\cite{Liu:15} use neural networks to solve the cardinality estimation problem, but focus on selection queries only. Hasan \ea~\cite{Hasan:19} also only focus on selectivity estimation, but show that deep learning models are particularly successful at predicting query cardinalities with a large number of selection predicates. 

Work by Kipf \ea~\cite{Kipf:19} proposes a new deep learning approach to cardinality estimation by using a multi-set convolutional network. Cardinality estimation does improve, but they do not show improvement of query plans. In addition, our work explores the space, time, accuracy of these models across a variety of datasets. Work by Woltmann \ea~\cite{Woltmann:19} propose building specialized neural network models that focus on a specific part of the schema (i.e. a join between two relations). These local models are beneficial as they reduce the query sample space required for training, and thus, reduce the training time. In our work, we propose to use active learning as an approach to help reduce the number of training samples. Work by Dutt \ea~\cite{Dutt:19} also compare tree ensembles and neural network models for cardinality estimation, but only focus on selectivities for single relations.

Work by Kraska \ea~\cite{Kraska:18} uses a mixture of neural networks to learn the distribution of an attribute with a focus on building fast indexes. In SageDB ~\cite{Kraska:19}, this work is extended towards building a new system that learns the underlying structure of the data to provide optimal query plans. In their work, they state that one key aspect in successfully improving these query plans is through cardinality estimation. They are currently working on a hybrid model-based approach to cardinality estimation, where they balance between looking for a model that can learn the distribution of the data and a model that can capture the extreme outliers and anomalies of the data. 

Wu \ea~\cite{Wu:18b} learn several models to predict the cardinalities for a variety of template subgraphs in a dataset instead of building one large model. Input features include filters and parameters for the subgraph, but they do not featurize information about the dataset (i.e. the relations). Thus, their models cannot make predictions for unobserved subgraph templates.

%% file: conclusion.tex
\section{Conclusion}\label{sec:discussion} 
We show the promise of using deep learning models to predict query cardinalities. In our study we found that even simple models can improve the runtimes of several queries across a variety of datasets. 


\textbf{Acknowledgements} This project was supported in part by the Graduate Opportunities and Minority Achievement Program (GO-MAP) fellowship, NSF grant IIS-1524535 and Teradata.